\documentclass[preprint,onecolumn,10pt,groupedaddress,showpacs,aps,floatfix,fleqn,amsmath]{revtex4}
\usepackage{amsfonts}

\usepackage{wrapfig} 
\usepackage[colorlinks]{hyperref} 
\usepackage[colorinlistoftodos, textwidth=4cm, shadow]{todonotes}
\usepackage{float}
\usepackage{subfig}
\usepackage{booktabs} 
\usepackage{amsmath}
\usepackage[normalem]{ulem} 

\graphicspath{Figs/}

\begin{document}
%
 
\title{Atomistic $k\cdot p$ theory}
\author{Craig E. Pryor}
\email{craig-pryor@uiowa.edu}
\affiliation{ Department of Physics and Astronomy, University of Iowa, Iowa City, Iowa, 52242, USA}

\author{ M.-E. Pistol}
\email{mats-erik.pistol@ftf.lth.se} 
\affiliation{NanoLund and Solid State Physics, Lund University, P.O. Box 118, 221 00 Lund, Sweden}

\date{\today}
\begin{abstract}
Pseudopotentials, tight-binding models, and $k\cdot p$ theory have stood for many years as the standard techniques for computing electronic states in crystalline solids.
Here we present the first new method in decades, which we call atomistic $k\cdot p$ theory.
In its usual formulation, $k\cdot p$ theory has the advantage of depending on parameters that are directly related to experimentally measured quantities, however it is insensitive to the locations of individual atoms.
We construct an atomistic $k\cdot p$ theory  by defining envelope functions on a  grid matching the crystal lattice.
The model parameters are matrix elements which are obtained from experimental results or {\it ab initio} wave functions in a simple way. This is in contrast to the other atomistic approaches in which parameters are fit to reproduce a desired dispersion and are not expressible in terms of fundamental quantities. This fitting is often very difficult. 
We illustrate our method by constructing a four-band atomistic model for a diamond/zincblende crystal and show that it is equivalent to the $sp^3$ tight-binding model. We can thus directly derive the parameters in the $sp^3$ tight-binding model from experimental data.
We then take the atomistic limit of the widely used eight-band Kane model and compute the band structures for all III-V semiconductors not containing nitrogen or boron using parameters fit to experimental data.
Our new approach extends $k\cdot p$ theory to problems in which atomistic precision is required, such as impurities, alloys, polytypes, and interfaces. It also provides a new approach to multiscale modeling by allowing continuum and atomistic $k\cdot p$ models to be combined in the same system.
\end{abstract}
\pacs{
71.20.Nr,	
71.15.-m,	
71.15.Ap 
}

\maketitle

\section{Introduction}
\label{sect:introduction}

An electron in the periodic potential of a semiconductor may be described using pseudopotentials\cite{Cohen.pr.1966,Chelikowsky.prb.1976}, tight-binding models\cite{Vogl.jpcs.1983,Jancu.prb.1998,Klimeck.sm.2000b,Jancu.prb.2005}, or $k\cdot p$ theory\cite{Voon.book.2009}. 
All three have been applied to semiconductor nanostructures \cite{Wang.prb.1996b,Zunger.pss.2001,Williamson.prb.1999,Saito.prb.1998,Klimeck.sm.2000b,Jaskolski.prb.2006,Grundmann.prb.1995,Cusack.prb.1996,Jiang.prb.1997,Pryor.prb.1998} in which the translational symmetry is broken by heterostructures, an applied potential, or a finite size.
Pseudopotentials and tight-binding models are inherently atomistic in that they allow, and even require, specification of the locations of atoms.
In contrast, $k\cdot p$ theory provides a Hamiltonian for the coarse grained crystal using parameters that depend on the composition and structure of the material.
Alloys are treated in the virtual crystal approximation in which the parameters specifying the band structure are empirically fit to the observed electronic properties of a material.
While this precludes a description with atomic scale precision, typically one does not know the exact position of every atom in a system anyway.

The three methods involve tradeoffs in the approximations made and the physical phenomena that they describe most accurately.
While $k\cdot p$ theory is a continuum model, the momentum matrix elements which parameterize it depend on the atomic scale structure of the electronic wave functions. 
This is advantageous in the computation of optical properties since the dipole matrix elements depend on the momentum matrix elements of the Bloch functions which also determine the band structure.
Tight-binding models use atomistic scale wave functions but involve a large number of parameters which are determined using complicated fitting procedures\cite{Goodwin.epl.1989} such as genetic algorithms \cite{Klimeck.sm.2000a,Klimeck.sm.2000b}.
Pseudopotentials also require a large number of form factors and must rely on complex fitting procedures, especially when strain is involved\cite{Kim.prb.1998}.
Pseudopotentials are atomistic, but smooth out the core wave function which results in smaller momentum matrix elements and
thus smaller optical matrix elements.
By including enough bands, any of the three methods can be made to be accurate throughout the Brillouin zone
\cite{Cardona.pr.1966,Fraj.sst.2008,Fraj.jap.2007,Richard.prb.2005,Saidi.jap.2008}.
Here we will focus on the dispersion around zone center.

$k\cdot p$ theory in the envelope approximation has been used successfully to describe electronic states in a wide variety of inhomogeneous semiconductor systems. 
The electronic wave function is taken to be a sum of Bloch functions, each multiplied by a slowly varying envelope function.
The effective Hamiltonian for the envelopes consists of material-dependent coefficients multiplying derivatives acting on the envelopes.
The electronic states are then determined by putting the envelopes on a computational grid and using finite difference approximations for derivatives, giving a model that is coarse-grained over a size comparable to the grid spacing.

An interesting question arises: can the computational grid be made small enough to make the model atomistic?
Finite differences make the model superficially resemble a tight-binding model with hopping between grid sites, suggesting a connection between tight-binding and $k\cdot p$ models.
In this paper we will develop an atomistic $k\cdot p$ theory by constructing it on a grid in which the sites correspond to atomic positions in the crystal lattice.
We will show that the atomistic limit of a simple four-band $k\cdot p$ theory is equivalent to a tight-binding model, thus allowing the determination of $k\cdot p$ parameters and tight-binding parameters in terms of each other.
This connection may be used to derive atomistic models that identically reproduce the long wavelength physics of $k\cdot p$ theory. 

We will begin in Sect. \ref{sect:envelopeTheory} with a discussion of $k\cdot p$ theory in the envelope approximation (henceforth simply $k\cdot p$ theory) on a three-dimensional grid of points, and using finite differences.
We generalize this widely used method to an arbitrary grid which need not be cartesian or regular.
In Sect. \ref{sect:atomisticGrid} we take the grid to be the crystal lattice itself and introduce difference operators on a diamond or zincblende grid.
In Sect. \ref{sect:matrixElements} we discuss how matrix elements are computed on such an atomistic grid, and examine the new non-zero momentum matrix elements which appear.
In Sect. \ref{sect:four-bandModel} we take the atomistic limit of a simple four-band model without spin-orbit coupling and show that it is identical to a tight-binding model.
In Sect. \ref{sect:nonhermiticity} we examine the non-Hermiticity that can arise (as in the four-band model) and in Sect. \ref{sect:finiteVolumeMethod} show how it may be resolved using the finite volume method.
In Sect. \ref{sect:eight-bandModel} we take the atomistic limit of the more realistic (and widely used) eight-band Kane model with spin-orbit coupling.
In Sect. \ref{sect:parameterFitting} we describe our fitting method and present our numerical fits of the atomistic parameters for the non-nitride III-V semiconductors .
We conclude with a discussion of some of the unique features and merits of the atomistic limit.

 \section{envelope theory}
 \label{sect:envelopeTheory}
 We begin with the basics in order to establish notation.
The Hamiltonian for a single electron in a semiconductor is
\begin{subequations}
\begin{align}
\label{eq:H0}
\hat H &= \hat H_0 + \hat H_{so} \\ 
\hat H_0 &= \frac{\hat p^2}{2m_0}+V_0( \mathbf r)+V_e( \mathbf r) \\ 
\hat H_{so} &= \frac{\hbar}{4m_0^2c^2}\left( \mathbf{\sigma} \times \nabla V \right) \cdot \mathbf{\hat p} 
\end{align}
\end{subequations}
where $V_0( \mathbf{r})$ is the crystal potential and $V_e(\mathbf{r})$ is a possible externally applied potential.
In a translational invariant system the single electron states may be found using multi-band $k\cdot p$ theory by writing the wave function as
\begin{align}
\label{eq:planewave}
\psi_{\mathbf k}({\mathbf r}) = \sum_{n=1}^N a_{n \mathbf k} \exp( i {\mathbf k} \cdot {\mathbf r} ) u_n({\mathbf r})
\end{align}
where $ u_n({\mathbf r})$ are the zone-center Bloch functions and $a_{n \mathbf k}$ are numerical coefficients.
The Hamiltonian is then an $N\times N$ numerical matrix 
\begin{equation}
\begin{split}
 H_{mn} =
 -\frac{\hbar^2}{2m_0} k^2 ~\delta_{mn} 
 - \frac{i \hbar}{m_0} \mathbf k \cdot \langle u_m | \hat {\mathbf p} |u_n\rangle 
 + \langle u_m| \hat H_0 |u_n\rangle
 + \langle u_m| \frac{\hbar^2}{4m_0^2c^2}\left( \mathbf{\sigma} \times \nabla V \right) |u_n\rangle \cdot \mathbf{k}
\label{eq:firstOrderk.p}
\end{split}
\end{equation}
where $N$ is the number of zone-center Bloch functions included in the basis. 
We will initially omit the spin-orbit term, and restore it later in section \ref{sect:eight-bandModel}.
Note that there is no explicit requirement of small $\mathbf{k}$ and Eq. \ref{eq:firstOrderk.p} is exact except for errors introduced by truncating the basis.
For any finite value of $N$ the accuracy decreases with increasing $\mathbf k$ since the solution will not be accurately expressed as a finite sum of zone center Bloch functions.

If translation symmetry is broken due to a heterojunction or an applied potential then the plane waves of Eq. \ref{eq:planewave} are replaced with envelope functions, and the wave function becomes
\begin{align}
\psi({\mathbf r}) =\sum_{n=1}^{N} f_n({\mathbf r}) u_n({\mathbf r}) \label{eq:envelopePsi},
\end{align}
where the $u_n({\mathbf r})$ are Bloch functions, but otherwise quite arbitrary.
Substituting $\psi(\mathbf{r})$ into the Schr$\rm \ddot{o}$dinger equation, and multiplying both sides by $u^*_m(\mathbf{r})$ we obtain
\begin{subequations}
\begin{align}
&\bigg[ -\frac{\hbar^2}{2m_0} a_{mn}(\mathbf{r}) \nabla^2
 + \boldsymbol{P}_{mn}(\mathbf{r}) \cdot \nabla + v_{mn}(\mathbf{r}) \bigg] f_n(\mathbf{r}) = E a_{mn}(\mathbf{r}) f_n(\mathbf{r}) 
\label{eq:HenvelopeContinuum}\\
&a_{mn}(\mathbf{r}) = u_m^*(\mathbf{r}) u_n(\mathbf{r}) \label{eq:amn} \\ 
&\boldsymbol{P}_{mn}(\mathbf{r}) = \frac{\hbar}{i m_0}u_m^*(\mathbf{r}) \mathbf{\hat p} u_n(\mathbf{r}) \label{eq:pmn} \\ 
& v_{mn}(\mathbf{r}) = u_m^*(\mathbf{r}) \left[ \frac{1}{2 m_0} \mathbf{\hat p}^2 +V(\mathbf{r} ) \right] u_n(\mathbf{r})\label{eq:vmn}
\end{align}
\end{subequations}
where $\mathbf{P}_{mn}$ and $v_{mn}$ are simply functions (i.e. the operators they contain act only on the Bloch functions contained within them).
We may make one re-arrangement which will be useful when we consider Hermiticity,
\begin{subequations}
\begin{align}
\bigg[ -\frac{\hbar^2}{2m_0} \nabla a_{mn}(\mathbf{r}) \nabla
 + \boldsymbol{\mathcal{P}}_{mn}(\mathbf{r}) \cdot \nabla + v_{mn}(\mathbf{r}) \bigg] f_n(\mathbf{r}) = E a_{mn}(\mathbf{r}) f_n(\mathbf{r}) 
\label{eq:HenvelopeContinuumAsym}\\
\boldsymbol{\mathcal{P}}_{mn}(\mathbf{r}) = \frac{1}{ 2}\left( \boldsymbol{P}_{mn} - \boldsymbol{P}_{nm}^* \right).\label{eq:pmnAsym} 
\end{align}
\end{subequations}
Since $u_m$ may be chosen as real, $\boldsymbol{P}_{mn}$ is real, and therefore $\boldsymbol{\mathcal{P}}_{mn} = - \boldsymbol{\mathcal{P}}_{nm}$.
The above equations give the effective Schr$\rm \ddot{o}$dinger equation for the envelope functions in which the Bloch functions appear as parameters.
Note that both are exact even with an incomplete set of Bloch functions since the envelopes are arbitrary. 
For example the trivial case of a single Bloch function $u_1(\mathbf{r})=1$ simply gives back the original Schr$\rm \ddot{o}$dinger equation, in which case the solution would consist of an envelope with variation over atomic scales.
If a sufficiently large Bloch basis is used, however, the wave function is well approximated with slowly varying envelopes.
Eq. \ref{eq:HenvelopeContinuum} will be approximate if the $f_n$ are constrained, as they are when defined on a grid that imposes a momentum cutoff.

\begin{figure}
\includegraphics[width=0.5\columnwidth]{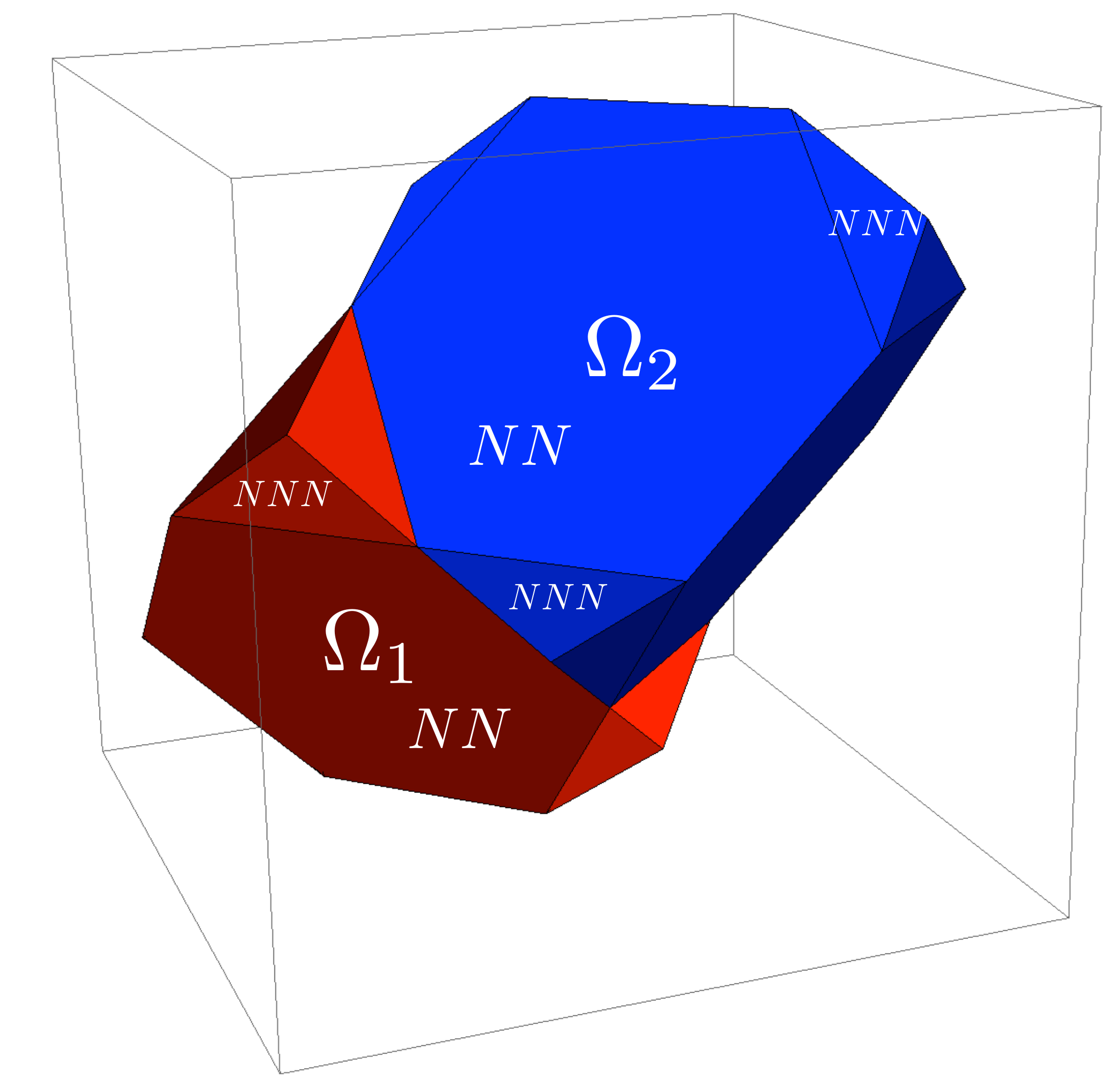}
\caption{Wigner Seitz cells in a zincblende crystal. $\Omega_1$ is the cell around the type-I atom at $(0,0,0)$ and $\Omega_2$ is the cell around the type-II atom at $(1/4,1/4,1/4)$. The hexagonal faces are the planes separating nearest neighbors and the small triangular faces are the planes separating second nearest neighbors, which are of the same type. If only the nearest neighbor planes were included, the cells would be tetrahedra. Taking next nearest neighbors into consideration truncates the corners of the tetrahedra, replacing them with (shorter) triangular pyramids.}
\label{fig:WSCell}
\end{figure}

 To obtain numerical solutions for systems that are not amenable to analytic methods requires reduction to a discrete system, such as by using a finite basis set or functions defined on a grid.
For nanostructures with irregular geometries, putting the envelope functions on a grid and replacing derivatives with finite differences is especially convenient since no assumptions about the geometric symmetry are required\cite{Pryor.prb.1991,Grundmann.prb.1995,Pryor.prb.1997}.
We denote the grid points with coordinates $\mathbf R$ and use $\mathbf{r}$ for continuum coordinates.
The continuous space can be broken up into cells $\Omega_\mathbf{R}$, each centered on the grid site at $\mathbf{R}$,
and the integral over all space can then be written as a sum of integrals over the cells
\begin{align}
\int d^3r = \sum_{\mathbf{R}} \int_{\Omega_{\mathbf{R}}} d^3 r.
\end{align}
If a sufficiently large number of Bloch functions are used the envelope functions will be slowly varying and $f_n(\mathbf{r})$ and its derivatives will be approximately constant over each cell.
This seemingly reasonable assumption can result in a non-Hermitian Hamiltonian, which will be resolved in Sect. \ref{sect:nonhermiticity}.
Integrating Eq. \ref{eq:HenvelopeContinuum} over $\Omega_\mathbf{R}$ and approximating derivatives of $f_n(\mathbf{r} )$ by finite differences on the grid we obtain
\begin{equation}
\bigg[ -\frac{\hbar^2}{2m_0} \langle u_m|u_n\rangle_{\Omega_{\mathbf R}} \Delta^2 
 + \frac{\hbar}{i m_0}
 \langle u_m| \mathbf{\hat p} |u_n\rangle_{\Omega_{\mathbf R}} \cdot \mathbf{\Delta} 
 + \langle u_m| \hat H_0 |u_n\rangle_{\Omega_{\mathbf R}} \bigg] f_{n\mathbf{R}}
= E \langle u_m|u_n\rangle_{\Omega_{\mathbf R}} f_{n\mathbf{R}}
\label{eq:HenvelopeDiscrete}\\ 
\end{equation}
where $f_{n\mathbf{R}}$ is the $n$th envelope function on the site at $\mathbf{R}$, and $\mathbf{\Delta}$ is the finite difference approximation to the gradient, $\mathbf{\Delta} f_{n\mathbf{R}} \approx \partial_x f_n(\mathbf{r})\big|_{\mathbf{r}=\mathbf{R}}$, which is a weighted sum of the values of $f_n$ at $\mathbf{R}$ and nearby grid sites.
We adopt an abbreviated notation for the projected matrix element of an operator $\mathcal{\hat O}$,
\begin{align}
 \int_{\Omega_\mathbf{R}} u_m^*(\mathbf{r}) \mathcal{\hat O} u_n(\mathbf{r}) ~d^3r =\langle u_m | \mathcal{\hat O} | u_n \rangle_{\Omega_\mathbf{R}} \label{eq:Pmn}.
 \end{align}
If $\Omega_\mathbf{R}$ contains an integer number of crystal unit cells then $ \langle u_m|u_n\rangle_{\Omega_{\mathbf R}} = \delta_{mn}$.
The solution of Eq. \ref{eq:HenvelopeDiscrete} is obtained by computing the eigenvalues and eigenvectors of a large sparse matrix, for which there are efficient algorithms\cite{Cullum.book.1985, Wang.jcp.1994}.

If the Bloch functions are the same throughout the structure then the $ \langle u_m| \mathbf{\hat p} |u_n\rangle_{\Omega_{\mathbf R}} $ are constants and Eq. \ref{eq:HenvelopeDiscrete} can be used directly to determine the electronic states.
This will be the case if confinement is provided by an externally applied potential or for a nanocrystal in which the vacuum is modeled as a large potential barrier.
In a heterostructure, however, the matrix elements appearing in Eq. \ref{eq:HenvelopeDiscrete} will vary spatially.
In the atomistic limit, in which the $\mathbf{R}$s correspond to individual atoms, the matrix elements will vary spatially even in a bulk crystal if it contains different atoms.
This will cause the Hamiltonian in Eq. \ref{eq:HenvelopeDiscrete} to be non-Hermitian, requiring a more careful treatment.
We will return to this problem and its remedy in section \ref{sect:nonhermiticity}.

\section{Atomistic grid}
 \label{sect:atomisticGrid}

 The finite difference approximation consists of replacing a derivative at a grid site with a difference operator which acts by taking weighted sums of the values on nearby grid sites.
For example, on a uniform cartesian grid the derivative of a function $f$ at a grid site located at $\mathbf{R}$ may be approximated using the symmetric difference
\begin{align}
\Delta_x f(\mathbf{r}) \bigg|_{\mathbf{r}=\mathbf{R}}= \frac{f(\mathbf{R}+\epsilon \mathbf{\hat x})-f(\mathbf{R}-\epsilon \mathbf{\hat x})}{2\epsilon} = \partial_x f(\mathbf{r})\bigg|_{\mathbf{r}=\mathbf{R}} + \mathcal O (\epsilon) \label{eq:fd}
\end{align}
where $\epsilon$ is the grid spacing and $ \mathbf{\hat x}$ is a unit vector.
This gives the lowest order approximation to $\partial_x$, and more accurate results may be obtained by including more sites in the sum\cite{Abramowitz.book.1964}.
The accuracy of the finite difference approximation improves as the grid spacing decreases, making it tempting to shrink the grid spacing as much as possible.
The existence of a physical crystal lattice 
suggests using the crystal lattice itself as the computational grid.
The values of the envelope functions will then be defined on the atoms themselves, yielding an atomistic theory.
We will develop this model for the diamond/zincblende lattice because of its importance in semiconductor physics, but the approach can be applied to any crystal structure.

On a rectangular grid the low order finite difference approximations may be written down intuitively, but on a non-rectilinear grid one needs a more systematic approach.
The general method for constructing a difference operator is to write down a Taylor series expansion of a function at a point $\mathbf{R}$, express the function values on sites in terms of that expansion, and solve for the linear combination of the function values on the site and its neighbors that gives the desired derivative to lowest order\cite{Fornberg.mc.1988,Lakin.ijnme.1986,Abramowitz.book.1964,Beck.rmp.2000}.
This method is used to obtain high-order difference approximations with smaller errors, and it may be used with non-rectilinear grids or even irregular grids.

Because the zincblende/diamond lattice has a basis with two atoms per unit cell the difference operators will be different on the inequivalent sites.
We denote the atom at $(0,0,0)$ as type 1 (assumed to be the anion in zincblende) and the atom at $\frac{a_{latt}}{4}(1,1,1)$ as type 2 (cation in zincblende), where $a_{latt}$ is the lattice constant.
The nearest neighbors of the type 1 atoms are at displacements
\begin{subequations}
\begin{align}
{\mathbf d_1} &= \frac{a_{latt}}{4}(1,1,1)\\ 
{\mathbf d_2} &= \frac{a_{latt}}{4}(-1,-1,1)\\ 
{\mathbf d_3} &= \frac{a_{latt}}{4}(-1,1,-1)\\ 
{\mathbf d_4} &= \frac{a_{latt}}{4}(1,-1,-1)
\end{align}
\end{subequations}
and the nearest neighbors of the type 2 atoms are at displacements $-{\mathbf d_n}$.
On the type 1 sites the nearest neighbor differences are given by
\begin{subequations} \label{eq:diff1s}
\begin{align}
\partial_x f({\mathbf r})\big|_{\mathbf{r}=\mathbf{R}}
&= \Delta_{1x} f({\mathbf R}) + {\mathcal O}(a_{latt}) =\big[ f({\mathbf R}+{\mathbf d_1}) -f({\mathbf R}+{\mathbf d_2}) -f({\mathbf R}+{\mathbf d_3}) +f({\mathbf R}+{\mathbf d_4}) \big]/a_{latt} + {\mathcal O}(a_{latt}) \label{eq:dx1}\\ 
\partial_y f({\mathbf r})\big|_{\mathbf{r}=\mathbf{R}} 
&= \Delta_{1y} f({\mathbf R})+ {\mathcal O}(a_{latt}) =\big[ f({\mathbf R}+{\mathbf d_1}) -f({\mathbf R}+{\mathbf d_2}) +f({\mathbf R}+{\mathbf d_3}) -f({\mathbf R}+{\mathbf d_4}) \big]/a_{latt} + {\mathcal O}(a_{latt}) \\ 
\partial_z f({\mathbf r})\big|_{\mathbf{r}=\mathbf{R}} 
&= \Delta_{1z} f({\mathbf R})+ {\mathcal O}(a_{latt}) =\big[ f({\mathbf R}+{\mathbf d_1}) +f({\mathbf R}+{\mathbf d_2}) -f({\mathbf R}+{\mathbf d_3}) -f({\mathbf R}+{\mathbf d_4}) \big]/a_{latt} + {\mathcal O}(a_{latt}) \\ 
\nabla^2 f({\mathbf r})\big|_{\mathbf{r}=\mathbf{R}} 
&= \Delta_1^2 f({\mathbf R})+ {\mathcal O}(a_{latt}) = \frac{8}{a_{latt}^2}\big[ f({\mathbf R}+{\mathbf d_1}) +f({\mathbf R}+{\mathbf d_2}) +f({\mathbf R}+{\mathbf d_3}) +f({\mathbf R}+{\mathbf d_4}) -4f({\mathbf R})\big] + {\mathcal O}(a_{latt}) 
\end{align}
\end{subequations}
and on the type 2 sites the difference operators are
\begin{subequations} \label{eq:diff2s}
\begin{align}
\Delta_{2x} f({\mathbf R}) &=\big[-f({\mathbf R}-{\mathbf d_1}) +f({\mathbf R}-{\mathbf d_2}) +f({\mathbf R}-{\mathbf d_3}) -f({\mathbf R}-{\mathbf d_4}) \big]/a_{latt} \label{eq:dx2}\\ 
\Delta_{2y} f({\mathbf R}) &=\big[ -f({\mathbf R}-{\mathbf d_1}) +f({\mathbf R}-{\mathbf d_2}) -f({\mathbf R}-{\mathbf d_3}) +f({\mathbf R}-{\mathbf d_4}) \big]/a_{latt} \label{eq:dy2}\\
\Delta_{2z} f({\mathbf R}) &=\big[ -f({\mathbf R}-{\mathbf d_1}) -f({\mathbf R}-{\mathbf d_2}) +f({\mathbf R}-{\mathbf d_3}) +f({\mathbf R}-{\mathbf d_4}) \big]/a_{latt} \label{eq:dz2}\\ 
\Delta_2^2 f({\mathbf R}) &=\frac{8}{a_{latt}^2}\big[ f({\mathbf R}-{\mathbf d_1}) +f({\mathbf R}-{\mathbf d_2}) +f({\mathbf R}-{\mathbf d_3}) +f({\mathbf R}-{\mathbf d_4}) -4f({\mathbf R})\big]. \label{eq:d22}
\end{align}
\end{subequations}
Using the site at $\mathbf{R}$ and its four nearest neighbors, only the four derivatives $\mathbf \nabla$ and $\nabla^2$ can be constructed. 
Substituting the above difference operators in Eq. \ref{eq:HenvelopeDiscrete} gives the Hamiltonian for the envelopes, parameterized by the matrix elements in Eq.s \ref{eq:HenvelopeDiscrete} which would be empirically fit to measurements on bulk materials.

Perturbative $k\cdot p$ models require approximations for second derivatives as well.
Since no derivative approximation beyond $\mathbf \nabla$ and $\nabla^2$ can be constructed with four nearest neighbors, second nearest neighbors must be used.
 The second nearest neighbor differences are the same for type 1 and 2 sites,
 \begin{subequations}
\begin{align}
\partial_x^2 f(\mathbf{R}) = \frac{1}{a_{latt}^2}\bigg[ 
 4f(\mathbf{R})
-f(\mathbf{R}+\mathbf{d}_{-1,-1,0} ) 
-f(\mathbf{R}+\mathbf{d}_{1,-1,0} ) 
-f(\mathbf{R}+\mathbf{d}_{-1, 1,0} ) 
-f(\mathbf{R}+\mathbf{d}_{1, 1,0} ) \nonumber \\
-f(\mathbf{R}+\mathbf{d}_{-1,0,-1} ) 
-f(\mathbf{R}+\mathbf{d}_{1,0,-1} ) 
-f(\mathbf{R}+\mathbf{d}_{-1,0, 1} ) 
-f(\mathbf{R}+\mathbf{d}_{1,0, 1} ) \nonumber \\
+f(\mathbf{R}+\mathbf{d}_{0,-1,-1} ) 
+f(\mathbf{R}+\mathbf{d}_{0,1,-1} ) 
+f(\mathbf{R}+\mathbf{d}_{0,-1, 1} ) 
+f(\mathbf{R}+\mathbf{d}_{0, 1, 1} )\bigg] \nonumber \\
+\mathcal{O}(a_{latt}^2)
\\
\partial_x \partial_y f(\mathbf x)\big|_{\mathbf{x}=\mathbf{R}} 
=\frac{1}{a_{latt}^2} \bigg[
 f(\mathbf{R}+\mathbf{d}_{1,-1,0} ) 
+f(\mathbf{R}+\mathbf{d}_{-1,1,0} ) 
- f(\mathbf{R}+\mathbf{d}_{1,1,0} ) 
- f(\mathbf{R}+\mathbf{d}_{-1,-1,0} ) 
\bigg]
+\mathcal{O}(a_{latt}^2)
\end{align}
\end{subequations}
where $\mathbf{d}_{i,j,k} = \frac{a_{latt}}{2} \left[ i \mathbf{\hat x} +j \mathbf{\hat y} +k \mathbf{\hat z} \right]$ is the displacement to the second nearest neighbor from the central site at $\mathbf{R}$.
Difference formulas for $\partial_y^2$, $\partial_z^2$, $\partial_x\partial_x$, and $\partial_y\partial_z$ are obtained by cyclic permutation of $x,y,z$.
Note that while the approximation to $\nabla^2$ involves only nearest neighbors, mixed derivatives and $\partial_x^2$, $\partial_y^2$, and $\partial_z^2$ require next nearest neighbors.
This does not present any fundamental or technical problems and is consistent with the source of these terms, second order perturbation theory.
Since the zeroth-order Hamiltonian contains nearest neighbor couplings from $\nabla^2$, the second order Hamiltonian contains next nearest neighbor couplings.

\section{matrix elements}
\label{sect:matrixElements}
The atomistic matrix elements are somewhat different from those usually appearing in $k\cdot p$ theory due to the projection of the Bloch states to atomistic cells.
 The most obvious choice for $\Omega_\mathbf{R}$ would be the Wigner Seitz cell around each $\mathbf{R}$, as shown in Fig. \ref{fig:WSCell} for the diamond/zincblende lattice.
 Many of the same selection rules from $k\cdot p$ theory apply since they depend on $T_d$ symmetry, which the atomic cells possess.
The most notable difference is the existence of matrix elements that are zero in the continuum model but are non-zero in the atomistic limit with cancellations between the atomistic cells.
Over the primitive unit cell, $\Omega_1 + \Omega_2$, the Bloch functions satisfy
\begin{align}
 \langle u_m | u_n \rangle_{\Omega_1+\Omega_2}= \langle u_m | u_n \rangle_{\Omega_1} + \langle u_m | u_n \rangle_{\Omega_2} = \delta_{mn}
\end{align}
but the atomistic matrix elements are not necessarily proportional to $\delta_{mn}$ since for $m\ne n$ there could be two nonzero terms that cancel.
If no two bands transform as the same representation of $T_d$, then $\langle u_m | u_n \rangle_{\Omega_{1,2}} \propto \delta_{mn}$. 
For example in a model with an $SXYZ$ basis, $\langle X | S \rangle_{\Omega_{1,2}} = 0$ because under a rotation by $\pi$ about the $y$-axis the crystal is invariant, but $\langle X | S \rangle_{\Omega_{1,2}}$ will change sign.
In this paper we will consider models in which all the states transform differently, and in particular models derived from an $SXYZ$ basis. 
In these cases the right hand side of Eq. \ref{eq:HenvelopeDiscrete} is a diagonal matrix, which we choose to be a multiple of the unit matrix, and there is no need to solve a generalized eigenvalue problem. It only necessary to multiply the left hand side by the inverse of this matrix.

The use of atomistic cells modifies the momentum
matrix elements as well.
For states projected to a volume $\Omega$, the momentum matrix element is given by
\begin{align}
 \int_{\Omega}u^*_m({\mathbf r} ) ~{\frac{\hbar}{i}\nabla}~ u_n({\mathbf r} ) ~d^3r 
 &= 
 \left( \int_{\Omega} u^*_n({\mathbf r} )~ {\frac{\hbar}{i}\nabla}~ u_m({\mathbf r} ) ~d^3r \right)^*
 + \int_{\Omega} ~ {\frac{\hbar}{i}\nabla}\left[ u^*_m({\mathbf r} ) u_n({\mathbf r} ) \right] ~d^3r \label{eq:byParts}
\end{align}
where the second integral on the right hand side will vanish due to periodicity if $\Omega$ contains an integer number of unit cells.
If $\Omega$ contains some fraction of a unit cell, then $ u^*_m({\mathbf r} ) u_n({\mathbf r} )$ is not periodic over $\Omega$ and there is no reason for the second integral on the right hand side to vanish.
We write the projected matrix element in the more compact form
\begin{align}
 \langle u_m | \mathbf{\hat p} | u_n \rangle_{\Omega_{i}}
= \langle u_n | \mathbf{\hat p} | u_m \rangle_{\Omega_{i}} + \Pi_{\Omega_i n m}\label{eq:nonselfadjoint}
\end{align}
where $\Pi_{\Omega_i n m}$ is given by the second integral on the right hand side of Eq. \ref{eq:byParts} and is nonzero only if $\Omega$ does not contain an integer number of unit cells.
If we sum over two sub-cells $\Omega_1$ and $\Omega_2$ to make a whole unit cell the correction must vanish, and therefore $\Pi_{\Omega_1 n m}=-\Pi_{\Omega_2 n m}$.

The atomistic momentum matrix elements between the conduction and valence bands obey the same selection rules as in continuum $k\cdot p$ theory since they rely only on $T_d$ symmetry, however because the projected matrix elements on the two atoms can be different we have
\begin{align}
iP_0 = \frac{\hbar}{m_0}\langle S | \hat p_x | X \rangle_{\Omega_1+\Omega_2} = \frac{\hbar}{m_0} \langle S | \hat p_x | X \rangle_{\Omega_1} + \frac{\hbar}{m_0} \langle S | \hat p_x | X \rangle_{\Omega_2} = iP_{a_1} + iP_{a_2}
\end{align}
where the subscript $a$ denotes that the matrix elements are projected to single atoms.
Throughout this paper we will use an $a$ subscript to distinguish atomistic parameters from those of the continuum $k\cdot p$ theory.
For the diamond crystal $P_{a_1}= P_{a_2}$ due to inversion symmetry, while for zincblende $P_{a_1}\ne P_{a_2}$.

In $k\cdot p$ models with more than one p-like band, such as the 16 and 14-band models\cite{Cardona.prb.1988,Pfeffer.prb.1996}, there are also matrix elements of the form
$iQ = \frac{\hbar}{m_0} \langle X_v | \hat p_y | Z_c \rangle$ 
where the $v$ and $c$ subscripts indicate the valence and conduction bands. 
Matrix elements with this general form but within the same band, such as $\langle X_v | \hat p_y | Z_v \rangle$, are zero by a simple symmetry argument which depends on the periodicity of the unit cell.
Due to invariance under a $180^\circ$ rotation about the y-axis,
 $\langle X_v | \hat p_y | Z_v \rangle = 
 \langle Z_v | \hat p_y | X_v \rangle$.
 For matrix elements over a whole unit cell we also have
 $\langle X_v | \hat p_y | Z_v \rangle = 
 \langle Z_v | \hat p_y | X_v \rangle^*$,
 and therefore
 $\langle X_v | \hat p_y | Z_v \rangle = 
 \langle X_v | \hat p_y | Z_v \rangle^*$.
 But since the states $|X\rangle$, $|Y\rangle$, $|Z\rangle$ can all be taken as real,
 $\langle X_v | \hat p_y | Z_v \rangle$ must be pure imaginary and
therefore the matrix element is zero.
In the atomistic case, we see from Eq. \ref{eq:nonselfadjoint} that 
$\langle X_v | \hat p_y | Z_v \rangle_{\Omega_i}  \ne \langle X_v | \hat p_y | Z_v \rangle_{\Omega_i} ^*$ and the above argument breaks down.
While the projected matrix element can be nonzero, the matrix element over the whole cell is zero and therefore
\begin{align}
0= \frac{\hbar}{m_0} \langle X_v | \hat p_y | Z_v \rangle_{\Omega_1} + \frac{\hbar}{m_0} \langle X_v | \hat p_y | Z_v \rangle_{\Omega_2} = iQ_{a} - iQ_{a}.
\end{align}
We see that the atomistic $k\cdot p$ model has an additional parameter not present in the continuum $k\cdot p$ model from which it is derived. 
In an inversion non-symmetric crystal the atomistic limit will also double the number of Hamiltonian matrix elements since there will be different matrix elements on each atom.

Summarizing, in the atomistic model we have
\begin{subequations}
\begin{align}
iP_{a_i}&= \frac{\hbar}{m_0}\langle S | \hat p_x | X \rangle_{\Omega_i}= \frac{\hbar}{m_0}\langle S | \hat p_y | Y \rangle_{\Omega_i}= \frac{\hbar}{m_0}\langle S | \hat p_z | Z \rangle_{\Omega_i}\\
iP'_{a_i}&= -\frac{\hbar}{m_0}\langle X | \hat p_x | S \rangle_{\Omega_i}= -\frac{\hbar}{m_0}\langle Y | \hat p_y | S \rangle_{\Omega_i}= -\frac{\hbar}{m_0}\langle Z | \hat p_z | S \rangle_{\Omega_i} \\
 iP_{a_1}+iP_{a_2} &= -iP'_{a_1}-iP'_{a_2} = iP_0\label{eq:Pa}
\end{align}
\end{subequations}
 where $P_0$ is the usual continuum $k\cdot p$ parameter.
 In addition, there are new intraband matrix elements
 \begin{align}
iQ_{a_i} &= \frac{\hbar}{m_0} \langle X |\hat p_y |Z \rangle_{\Omega_i} = \frac{\hbar}{m_0} \langle Z |\hat p_x |Y \rangle_{\Omega_i} = \frac{\hbar}{m_0} \langle Y |\hat p_z |X \rangle_{\Omega_i}\nonumber \\ 
&= \frac{\hbar}{m_0} \langle Z |\hat p_y |X \rangle_{\Omega_i} = \frac{\hbar}{m_0} \langle X |\hat p_z |Y \rangle_{\Omega_i} = \frac{\hbar}{m_0} \langle Y |\hat p_x |Z \rangle_{\Omega_i} \label{eq:Qa}
\end{align}
which satisfy
$ iQ_{a_1} =-iQ_{a_2}$.

As will be seen in Sec. \ref{sect:four-bandModel}, in an inversion non-symmetric crystal the combination of finite differences and the fact that $P_i \ne P'_i$ results in a non-Hermitian Hamiltonian.
One solution is to simply use an inversion symmetric basis, which is reasonable since 
$k\cdot p$ theory is often formulated in the symmetric approximation.
As will be shown in Sect. \ref{sect:eight-bandModel}, inversion symmetry may still be broken by the sub-unit cell structure of the envelope function.
Alternatively, we may change the volumes of $\Omega_1$ and $\Omega_2$ by using generalized Voronoi cells\cite{Mishev.nmpde.1998,Telea.book.2011}.
By adjusting $\Omega_1$ and $\Omega_2$ we can make $P_i = P'_i$ while maintaining inversion non-symmetry.
A generalized Voronoi cell may be constructed by rescaling the distances that would be used to determine the Wigner Seitz cell.
Consider a site at $\mathbf R$, with nearest neighbors at $\mathbf R_{NN,i}$, and next nearest neighbors at $\mathbf R_{NNN,j}$.
The generalized Voronoi cell around $\mathbf R$ is the set of points $\mathbf p$ satisfying the two conditions
\begin{subequations}
\begin{align}
a& \big| \mathbf{p}-\mathbf{R} \big| < \big| \mathbf{p}-\mathbf R_{NN,i} \big|  \\
 &\big| \mathbf{p}-\mathbf{R} \big| < \big| \mathbf{p}-\mathbf R_{NNN,j} \big| 
\end{align}
\end{subequations}
where $a$ is a scaling factor that determines the relative sizes of a cell.
Note that the second condition does not have a scaling factor because the next nearest neighbors are of the same type.
Using $a\ne 1$ on the type-1 sites, and $a\rightarrow 1/a$ on the type-2 sites we change the relative volumes of $\Omega_1$ and $\Omega_2$ while maintaining $\Omega_1+ \Omega_2$ as a unit cell.
Shrinking $\Omega_1$ will decrease both $P_{a_1}$ and $P'_{a_1}$ while increasing $P_{a_2}$ and $P'_{a_2}$, and therefore we may adjust the scaling factor to make $P_{a_1}=P'_{a_2}$ and $P'_{a_1}=P_{a_2}$.
This method will prove useful for restoring Hermiticity in the inversion non-symmetric case in Sec. \ref{sect:four-bandModel}.

\section{four-band model}
\label{sect:four-bandModel}
To demonstrate the basic structure of the atomistic limit we first consider the simple four-band $k\cdot p$ model without spin orbit coupling, using zone-center Bloch states $|X\rangle$, $|Y\rangle$, $|Z\rangle$ for the valence band and $|S\rangle$ for the conduction band. 
This model and the tight-binding model with which it will be compared are too simple to be used for realistic calculations, but they demonstrate the basic structure of the atomistic limit.
Using plane waves for the envelopes, the $k\cdot p$ Hamiltonian is 
\begin{subequations}
\begin{equation}
H_0=
\bordermatrix{
~&S & X & Y & Z \cr
~&E_c + {\mathcal E_c}(k) & iP_0 k_x & iP_0 k_y & iP_0 k_z \cr
~& -iP_0 k_x & E_v+ {\mathcal E_v}(k)& 0 & 0 \cr
~& -iP_0 k_y & 0 & E_v+ {\mathcal E_v}(k) & 0 \cr
~& -iP_0 k_z & 0 & 0 & E_v+ {\mathcal E_v}(k)\cr
}\label{eq:H04band}
\end{equation}
\begin{align}
iP_0 &= \frac{\hbar}{m_0}\langle S | \hat p_x | X \rangle \\ 
 {\mathcal E_c}(k) &= \left(\frac{1}{2}+F_c \right) \frac{\hbar^2}{m_0} k^2 \\ 
 {\mathcal E_v}(k) &= \left(\frac{1}{2}+F_v \right)\frac{\hbar^2}{m_0} k^2 \label{eq:kp4}
\end{align}
\end{subequations}
where $F_c$ and $F_v$ are remote band contributions to the conduction and valence band respectively, and are included to make the $k\cdot p$ model agree with a tight-binding model.
In the atomistic limit the Hamiltonian also includes terms involving the momentum matrix element of Eq. \ref{eq:Qa}
\begin{subequations}
\begin{align}
H_Q=\bordermatrix{
~&S & X & Y & Z \cr
~& 0 & 0 & 0 & 0 \cr
~& 0 & 0 & i Q_a k_z & i Q_a k_y \cr
~& 0 & i Q_a k_z & 0 & i Q_a k_x \cr
~& 0 & i Q_a k_y & i Q_a k_x & 0 \cr
}\label{eq:HQ4band}\\
iQ_a = \frac{\hbar}{m_0}\langle X | \hat p_y | Z \rangle_{\Omega_1} = - \frac{\hbar}{m_0}\langle X | \hat p_y | Z \rangle_{\Omega_2}.
 \end{align}
 \end{subequations}
Eq. \ref{eq:HQ4band} is only for an anion site, and on the cation site we will have $-H_Q$.
 In a plane wave basis, it is convenient to define the approximate wave vectors obtained from the finite difference operators acting on plane waves,
\begin{subequations}
\begin{align}
\mathcal K_{1n} &= -i e^{-i {\mathbf k} \cdot {\mathbf r}} \Delta_{1n} ~e^{i {\mathbf k} \cdot {\mathbf r}} ~~~~~ (n = x,y,z)\\ 
\mathcal K_1^2 &= - e^{-i {\mathbf k} \cdot {\mathbf r}} \Delta_1^2 ~e^{i {\mathbf k} \cdot {\mathbf r}} \\ 
\mathcal K_{2n} &= -i e^{-i {\mathbf k} \cdot {\mathbf r}} \Delta_{2n} ~e^{i {\mathbf k} \cdot {\mathbf r}} = {\mathcal K_{1n}}^* ~~~~~ (n = x,y,z)\\ 
\mathcal K_2^2 &= - e^{-i {\mathbf k} \cdot {\mathbf r}} \Delta_2^2 ~e^{i {\mathbf k} \cdot {\mathbf r}} = {\mathcal K_1^2}^*
 \label{eq:approxks}
\end{align}
\end{subequations}
where the subscripts 1 and 2 on $\mathcal K$ indicates the atom type on which the difference is centered.
On a diamond/zincblende lattice, using Eq. \ref{eq:HenvelopeDiscrete} and the finite differences defined in Eq. \ref{eq:dx1}-\ref{eq:d22} we obtain
\begin{align}
H&=
\bordermatrix{
~ & S_1 & X_1 &Y_1 & Z_1 & S_2 & X_2 &Y_2 & Z_2\cr
~ & E_{c1}+ {\mathcal V_{c}} & 0 & 0 & 0 & {\mathcal T_{c1}} & iP_{a_1}~\mathcal K_{1x} & iP_{a_1}~\mathcal K_{1y} & iP_{a_1}~\mathcal K_{1z} \cr
~ & 0 & E_{v1}+{\mathcal V_{v}} & 0 & 0 & -iP'_{a_1}~\mathcal K_{1x} & {\mathcal T_{v1}} & iQ_a~\mathcal K_{1z} & iQ_a~\mathcal K_{1y} \cr
~ & 0 & 0 & E_{v1}+{\mathcal V_{v}} & 0 & -iP'_{a_1}~\mathcal K_{1y} & iQ_a~\mathcal K_{1z} & {\mathcal T_{v1}} & iQ_a~\mathcal K_{1x} \cr
~ & 0 & 0 & 0 & E_{v1}+{\mathcal V_{v}} & -iP'_{a_1}~\mathcal K_{1z} & iQ_a~\mathcal K_{1y} & iQ_a \mathcal K_{1x} & {\mathcal T_{v1}} \cr
~ &{\mathcal T_{c2}} & iP_{a_2}~\mathcal K_{2x} & iP_{a_2}~\mathcal K_{2y} & iP_{a_2}~\mathcal K_{2z} & E_{c2}+ {\mathcal V_{c}} & 0 & 0 & 0 \cr
~ & -iP'_{a_2}~\mathcal K_{2x} & {\mathcal T_{v2}} & -iQ_a \mathcal K_{2z} & -iQ_a \mathcal K_{2y} & 0 & E_{v2}+ {\mathcal V_{v}} & 0 & 0 \cr
~ & -iP'_{a_2}~\mathcal K_{2y} & -iQ_a~\mathcal K_{2z} & {\mathcal T_{v2}} & -iQ_a \mathcal K_{2x} & 0 & 0 & E_{v2}+ {\mathcal V_{v}} & 0 \cr
~ & -iP'_{a_2}~\mathcal K_{2z} & -iQ_a~\mathcal K_{2y} & -iQ_a~\mathcal K_{2x} & {\mathcal T_{v2}} & 0 & 0 & 0 & E_{v2}+ {\mathcal V_{v}} 
}\label{eq:atomistickp4} 
\end{align}
where the subscripts $1,2$ label the atoms within the unit cell and
\begin{subequations}
\begin{align}
\mathcal T_{bc}&= \left(\frac{1}{2}+F_b \right) \frac{\hbar^2}{m_0} \left( \mathcal K_a^2 -\frac{32}{a_{latt}^2} \right)\\ 
\mathcal V_{b}&= \left(\frac{1}{2}+F_b \right) \frac{\hbar^2}{m_0} \left( \frac{32}{a_{latt}^2} \right)
\end{align}
\end{subequations}
where $b$ is the band index ($v$ or $c$ for valence or conduction), $c$ indicates the type of atom (1 or 2), and the momentum matrix elements are given by Eq.s \ref{eq:Pa} and \ref{eq:Qa}.

 If the crystal lacks inversion symmetry, then $P_{a_1} \ne P'_{a_2}$, $P'_{a_1}\ne P_{a_2}$, and $H$ is not Hermitian.
 The problem may be remedied by starting with a set of inversion symmetric Bloch functions, in which case $P_{a_1}= P_{a_2} = P'_{a_1}=P'_{a_2}$.
 The Hamiltonian 
 can
 still be inversion non-symmetric due to the different potentials on the anion and cation, giving rise to inversion non-symmetric zone-center envelope functions.
An alternative approach is to modify the differencing scheme using the generalized Voronoi cells described at the end of Section \ref{sect:matrixElements}.
The cell size can be adjusted so as to make $P_{a_2}=P'_{a_1}$ and $P'_{a_2}=P_{a_1}$, restoring Hermiticity while maintaining $P_{a_1}\ne P'_{a_1}$ and $P_{a_2}\ne P'_{a_2}$.
Deforming the cells will also change the values of the $Q_{a_i}$, but since $Q_{a_1}=-Q_{a_2}$ the form of $H$ will not be affected.
In a model with more bands, modifying $\Omega$ to make $H$ Hermitian would appear to be limited to tuning the $P_{a_i}$s for just one pair of bands.
One could use different $\Omega$s for different bands, or simply set the $P_{a_i}$s on an {\it ad hoc} basis.

We may compare the atomistic four-band $k\cdot p$ model of Eq. \ref{eq:atomistickp4} with the four-band tight-binding Hamiltonian\cite{Chadi.pss.1975} 
\begin{equation}
H_{tb}=
\bordermatrix{
~ & S_1 &X_1 &Y_1 & Z_1 & S_2 & X_2 &Y_2 & Z_2 \cr
~ & E_{s1} & 0 & 0 & 0 & V_{ss} g_1 & V_{s p} g_2 & V_{s p} g_3 & V_{s p} g_4 \cr
~ & 0 & E_{p1} & 0 & 0 & -V_{s p} g_2 & V_{xx} g_1 & V_{xy} g_4 & V_{xy} g_3 \cr
~ & 0 & 0 & E_{p1} & 0 & -V_{s p} g_3 & V_{xy} g_4 & V_{xx} g_1 & V_{xy} g_2 \cr
~ & 0 & 0 & 0 & E_{p1} & -V_{s p} g_4 & V_{xy} g_3 & V_{xy} g_2 & V_{xx} g_1 \cr
~ & V_{ss}g_1^* & -V_{s p} g_2^* & -V_{s p}g_3^* & -V_{s p}g_4^* &E_{s2} & 0 & 0 & 0 \cr
~ & V_{s p}g_2^* & V_{xx} g_1^* & V_{xy} g_4^* & V_{xy} g_3^* & 0 & E_{p2} & 0 & 0 \cr
~ & V_{s p}g_3^* & V_{xy} g_4^* & V_{xx} g_1^* & V_{xy} g_2^* & 0 & 0 & E_{p2} & 0 \cr
~ & V_{s p}g_4^* & V_{xy} g_3^* & V_{xy} g_2^* & V_{xx} g_1^* & 0 & 0 & 0 & E_{p2} \cr
}\label{eq:tb}
\end{equation}
where the $g$s are the standard tight-binding functions, which are related to the $\mathcal K$s by
\begin{align}
g_1&=\frac{1}{4}\left[\exp(i {\mathbf d_1}\cdot {\mathbf k} )+\exp(i {\mathbf d_2}\cdot {\mathbf k} )+\exp(i {\mathbf d_3}\cdot {\mathbf k} )+\exp(i {\mathbf d_4}\cdot {\mathbf k} ) \right] = 1-\frac{a_{latt}^2}{32} \mathcal K^2\nonumber \\ 
g_2&=\frac{1}{4}\left[\exp(i {\mathbf d_1}\cdot {\mathbf k} )+\exp(i {\mathbf d_2}\cdot {\mathbf k} )-\exp(i {\mathbf d_3}\cdot {\mathbf k} )-\exp(i {\mathbf d_4}\cdot {\mathbf k} ) \right] = \frac{ia_{latt}}{4} \mathcal K_{x}\nonumber \\ 
g_3&=\frac{1}{4}\left[\exp(i {\mathbf d_1}\cdot {\mathbf k} )-\exp(i {\mathbf d_2}\cdot {\mathbf k} )+\exp(i {\mathbf d_3}\cdot {\mathbf k} )-\exp(i {\mathbf d_4}\cdot {\mathbf k} ) \right] = \frac{ia_{latt}}{4}\mathcal K_{y}\\ 
g_4&=\frac{1}{4}\left[\exp(i {\mathbf d_1}\cdot {\mathbf k} )-\exp(i {\mathbf d_2}\cdot {\mathbf k} )-\exp(i {\mathbf d_3}\cdot {\mathbf k} )+\exp(i {\mathbf d_4}\cdot {\mathbf k} ) \right] = \frac{ia_{latt}}{4} \mathcal K_{z} \nonumber
\end{align}
Equating the matrix elements in Eqs \ref{eq:atomistickp4} and \ref{eq:tb}, the tight-binding and $k\cdot p$ parameters are related by
\begin{subequations}
\begin{align}
V_{ss}=-\frac{\hbar^2}{m_0} \frac{32}{a_{latt}^2} \left( \frac{1}{2}+F_c \right)\\ 
V_{xx}=-\frac{\hbar^2}{m_0} \frac{32}{a_{latt}^2} \left( \frac{1}{2}+F_v \right)\\ 
V_{s_1p_2}=4 P_{a_1} / a_{latt}\\ 
V_{s_2p_1}=4 P_{a_2} / a_{latt}\\ 
V_{xy}=4 Q_a / a_{latt}\\ 
E_{s1}=E_{c1}+\frac{\hbar^2}{m_0} \frac{32}{a_{latt}^2}\left( \frac{1}{2}+F_c \right)\\ 
E_{p1}=E_{v1}+\frac{\hbar^2}{m_0} \frac{32}{a_{latt}^2}\left( \frac{1}{2}+F_v \right)\\ 
E_{s2}=E_{c2}+\frac{\hbar^2}{m_0} \frac{32}{a_{latt}^2}\left( \frac{1}{2}+F_c \right)\\ 
E_{p2}=E_{v2}+\frac{\hbar^2}{m_0} \frac{32}{a_{latt}^2}\left( \frac{1}{2}+F_v \right).
\end{align}
\end{subequations}
We thus find that the atomistic limit of the $k\cdot p$ model is equivalent to a tight-binding model.
In order to make this one-to-one correspondence it was necessary to include spherically symmetric remote band contributions to both the conduction and valence bands and to have different momentum matrix elements on the two atoms (at least in the inversion non-symmetric case).
Our inclusion of only limited remote band contributions to the valence band gives the Luttinger model in the spherical approximation with $\bar \gamma = \frac{1}{5} \left( 2\gamma_2 + 3\gamma_3 \right) $.

The atomistic limit always adds at least one new parameter, $Q_a$. 
For inversion symmetric crystals the Hamiltonian matrix elements are the same on both atoms, so there are as many parameters as in the original $k\cdot p$ model, plus the additional parameter $Q_a$.
For inversion non-symmetric crystals the Hamiltonian matrix elements are different on each atom, so the number of diagonal matrix elements is doubled, plus the additional $Q_a$.
In Sect. \ref{sect:eight-bandModel} we will take a hybrid approach in which the Hamiltonian is not inversion symmetric, but the Bloch basis is chosen to be symmetric. 
In that case the matrix elements of $\hat H_0$ (c.f. Eq. \ref{eq:H0}) will be different on different atoms, but the momentum matrix elements will be the same on each atom since they depend on derivatives of the inversion symmetric Bloch functions.
Therefore, the number of diagonal parameters will be doubled, the number of momentum matrix elements will remain the same, and $Q_a$ will be added.

An important feature of the atomistic limit is that the envelope varies within a unit cell even at zone center, and thus modifies the effective Bloch functions.
In Eq. \ref{eq:atomistickp4} we see that $\Delta^2$ couples the two atoms even at $\mathbf k = 0$ via the off-diagonal matrix elements ${\mathcal T}_{c1}$, ${\mathcal T}_{v1}$, ${\mathcal T}_{c2}$, and ${\mathcal T}_{v2}$.
This results in a doubling of the number of bands over the continuum model, with the additional bands being shifted by an energy on the order of $ \hbar^2 / m_0 a_{latt}^2$.
Since the atomistic model includes two grid sites per unit cell the envelope functions include wave vectors outside the first Brillouin zone. 
These states may be interpreted as approximate Bloch functions with different symmetry from the zone-center Bloch functions of the theory. 
For example, when multiplied by an envelope that changes sign from site to site the anti-bonding S-like Bloch function of the conduction band becomes similar to the bonding S-like state. Of course such a "fake" Bloch function is not the true zone-center Bloch function, but an approximation.
This simple model illustrates the basic features of the atomistic limit, but in order to develop more realistic models we need to examine the inversion non-symmetric case more closely and study the relationship to heterojunctions.

\section{Non Hermiticity}
\label{sect:nonhermiticity}
As we saw in Sec. \ref{sect:four-bandModel}, straight-forward application of the atomistic limit to an inversion non-symmetric crystal gives a non-Hermitian Hamiltonian since the momentum matrix elements are different on different atoms.
This problem generally arises when a finite difference operator is multiplied by a spatially varying coefficient, and also occurs at a heterojunction in continuum $k\cdot p$ theory\cite{Zhu.prb.1983,Pryor.prb.1991,Grundmann.prb.1995,Pryor.prb.1997}.
In this section and Sec. \ref{sect:finiteVolumeMethod} we will examine this issue in a general framework that is applicable to both continuum $k\cdot p$ models that have been put on a grid and our atomistic model.
Consider a Hamiltonian containing a Hermitian differential operator $\mathcal{D}$ which is approximated by a difference operator $\mathbb{D}$ consisting of a weighted sum of the function values on nearby grid sites,
\begin{align}
\mathcal{D} f(\mathbf{r})\bigg|_{\mathbf{r}=\mathbf{R}} \approx \mathbb{D} f(\mathbf{r})\bigg|_{\mathbf{r}=\mathbf{R}} = \sum_{\mathbf{R}'} d_{\mathbf{R} \mathbf{R}'} f_\mathbf{R'}
\end{align}
where $d_{\mathbf{R},\mathbf{R}'}$ are coefficients defining the difference operator, most of which are zero except when $\mathbf{R}'$ and $\mathbf{R}$ are close to each other.
Since $\mathbb{D}$ is Hermitian,
$\langle g | \mathbb{D} f \rangle = \langle \mathbb{D}g | f \rangle$
and the corresponding difference operator must satisfy
\begin{align}
\sum_{\mathbf{R},\mathbf{R}'} g_{\mathbf{R}}^* ~d_{\mathbf{R},\mathbf{R}'}~ f_{\mathbf{R}'} = \sum_{\mathbf{R},\mathbf{R}'} d_{\mathbf{R},\mathbf{R}'}^* ~g_{\mathbf{R}'}^*~ f_{\mathbf{R}} = \sum_{\mathbf{R},\mathbf{R}'} d_{\mathbf{R}',\mathbf{R}}^* ~g_{\mathbf{R}}^* ~f_{\mathbf{R}'} 
\end{align}
where the sums on $\mathbf{R}$ and $\mathbf{R}'$ range over all lattice sites.
Therefore the Hermiticity of $ \mathbb{D}$ requires $d_{\mathbf{R},\mathbf{R}'} = d_{\mathbf{R}',\mathbf{R}}^*$. 
If the continuum Hamiltonian contains a differential operator multiplied by a spatially varying coefficient $c(\mathbf{r})$ then simply multiplying that operator by $c(\mathbf{R})$ gives 
\begin{align}
c(\mathbf{r}) \mathcal{D} f(\mathbf{r})\bigg|_{\mathbf{r}=\mathbf{R}} \rightarrow c(\mathbf{r}) \mathbb{D} f(\mathbf{r})\bigg|_{\mathbf{r}=\mathbf{R}} = \sum_{\mathbf{R}} c_{\mathbf{R}} ~d_{\mathbf{R},\mathbf{R}'} ~f_{\mathbf{R}'}.
\end{align}
 Since Hermiticity requires
$c_{\mathbf{R}} d_{\mathbf{R},\mathbf{R}'} = \left[ c_{\mathbf{R}'} d_{\mathbf{R}',\mathbf{R}} \right]^*$, any spatial variation in the magnitude of $c(\mathbf{r})$ spoils the Hermiticity.
This problem arises in a one-band model with a spatially varying effective mass as well as in multi-band envelope models with spatially varying parameters. 
This problem will occur in Eq. \ref{eq:HenvelopeDiscrete} for a heterostructure, but in the atomistic limit it will arise even for a bulk material if the atoms differ from one another. 

A common solution is to symmetrize over the connected sites\cite{Pryor.prb.1991,Grundmann.prb.1995,Pryor.prb.1997}, 
\begin{align}
c(\mathbf{r}) \mathcal{D} f(\mathbf{r})\bigg|_{\mathbf{r}=\mathbf{R}} \rightarrow \sum_{\mathbf{R}'} \frac{1}{2}\left[ c_{\mathbf{R}}+c_{\mathbf{R}'} \right] d_{\mathbf{R},\mathbf{R}'} f_{\mathbf{R}'}\label{avgOverSites}.
\end{align}
The symmeterization is applied to $\Delta_x$, $\Delta_y$, and $\Delta_z$, which are then used to construct other operators.
This resolution of the problem is not as {\it ad hoc} as it may seem since the first derivative is most naturally defined on a link between two sites.
For example on a one dimensional grid with spacing $\epsilon$ the difference between two adjacent sites gives an approximation to the derivative on the link connecting them, 
\begin{align}
\partial_x f(x)\big|_{x=x_0+\epsilon/2} \approx \Delta_x f(x)\big|_{x=x_0+\epsilon/2} = \frac{1}{\epsilon} \left( f(x_0+\epsilon) - f(x_0) \right).
\end{align}
Therefore, the value of a coefficient multiplying the derivative is naturally defined on the link itself and should be interpolated between the two points being differenced, giving
\begin{align}
c(x) \partial_x f(x)\big|_{x=x_0+\epsilon/2} \approx \frac{1}{2\epsilon}\left[ c(x_0)+c(x_0+\epsilon) \right] \left[ f(x_0+\epsilon) - f(x_0) \right].
\end{align}
For terms containing variable coefficients and second derivatives, one must also be careful about the operator ordering.
Neither $c(x) \partial_x^2$ nor $\partial_x^2 c(x)$ can be made self-adjoint, even in the continuum, however a symmetrized operator such as $\partial_x c(x) \partial_x$ can \cite{Zhu.prb.1983,Birman.book.1986,Einevoll.prb.1990}.
Therefore we can write
\begin{align}
\partial_x c(x) \partial_x f(x)\big|_{x=x_0} &\approx
 \frac{1}{\epsilon} \left[ c_x \Delta_x f_x \big|_{x=x_0+\epsilon/2} - c_x\Delta_x f_x\big|_{x=x_0-\epsilon/2} \right] \nonumber \\ \relax
 &\approx \frac{1}{2\epsilon^2} \bigg[ (c_{x_0}+c_{x_0+\epsilon} ) f_{x_0+\epsilon} 
 + (c_{x_0}+c_{x_0- \epsilon} ) f_{x_0-\epsilon} 
 - (c_{x_0+\epsilon}+c_{x_0 -\epsilon}+2 c_{x_0}) f_{x_0} \bigg].
\end{align}

\section{Finite Volume Method}
\label{sect:finiteVolumeMethod}
The symmetrization procedure described above has an intuitive appeal, however a more formal approach will give the same result while providing some additional insight into the problem.
The root cause of the non-Hermiticity is that in discretizing Eq. \ref{eq:HenvelopeContinuum} we assumed the envelopes and their derivatives were constants over $\Omega_\mathbf{R}$.
Instead, we can make use of the finite volume method\cite{Knabner.book.2003} in which the divergence theorem is used to convert the volume integral over a cell into a surface integral.
Discretizing this modified version results in finite differences over two sites that are multiplied by a quantity defined on the link between the sites.
This means that the coefficient is the same (except for a possible sign) when evaluated on either of the sites, thus guaranteeing Hermiticity.

Let us return to the continuum, but using Eq. \ref{eq:HenvelopeContinuumAsym} rather than Eq. \ref{eq:HenvelopeContinuum}, and consider the $\mathbf{\mathcal{P}}_{mn}(\mathbf{r}) \cdot \nabla f_n(\mathbf{r})$ term. 
When integrated over a volume around the grid site $\mathbf{R}$,
\begin{align}
\int_{\Omega_\mathbf{R}}\mathbf{\mathcal{P}}_{mn}(\mathbf{r}) \cdot \nabla f_n(\mathbf{r}) ~d^3r
&=\int_{\partial \Omega_\mathbf{R}} f_n(\mathbf{r}) \mathbf{\mathcal{P}}_{mn}(\mathbf{r})\cdot d\mathbf{s} -\int_{\Omega_\mathbf{R}} f_n(\mathbf{r}) \nabla \cdot \mathbf{\mathcal{P}}_{mn}(\mathbf{r}) ~d^3r
\end{align}
where $\partial \Omega_\mathbf{R}$ is the bounding surface of $ \Omega_\mathbf{R}$.
Since $f_n(\mathbf{r})$ is slowly varying we can replace it with $f_{n\,\mathbf{R}}$ in the second integral on the right to obtain
\begin{align}
\int_{\Omega_\mathbf{R}}\mathbf{\mathcal{P}}_{mn}(\mathbf{r}) \cdot \nabla f_n(\mathbf{r}) ~d^3r &\approx \int_{\partial \Omega_\mathbf{R}}\ f_n(\mathbf{r}) \mathbf{\mathcal{P}}_{mn}(\mathbf{r}) \cdot d\mathbf{s} - f_{n\,\mathbf{R}} \int_{\partial \Omega_\mathbf{R}} \mathbf{\mathcal{P}}_{mn}(\mathbf{r}) \cdot d\mathbf{s}. \label{fvm1}
\end{align}
The surface $\partial \Omega_\mathbf{R}$ is a polyhedron centered on the point $\mathbf R$ with faces $S_\mathbf{d}$, each of which is normal to the displacement from $\mathbf{R}$ to the neighboring site at $\mathbf{R}+ \mathbf{d}$ (see Fig. \ref{fig:WSCell}).
Because $f$ is slowly varying, its value on one of the faces $S_\mathbf{d}$ is approximately constant, with a value that may be approximated by linearly interpolating between the two sites, $f_n(\mathbf{r})\big|_{S_\mathbf{d}} \approx \left[ f_{n\,\mathbf{R}} + f_{n\,\mathbf{R}+\mathbf{d}} \right]/2$.
This gives
\begin{subequations}
\begin{align}
\int_{\Omega_\mathbf{R}}\mathbf{\mathcal{P}}_{mn}(\mathbf{r}) \cdot \nabla f_n(\mathbf{r}) ~d^3r &\approx
\frac{1}{2} \sum_\mathbf{d} \left[ f_{n\,{\mathbf{R}+\mathbf{d}}} - f_{n\,\mathbf{R}} \right] \int_{S_\mathbf{d}} \mathbf{\mathcal{P}}_{mn}(\mathbf{r}) \cdot d\mathbf{s}\\\ 
&\approx
\frac{1}{2} \sum_\mathbf{d} P_{\mathbf{R} \mathbf{d}} ~\Delta_\mathbf{d} \,f_{n\,\mathbf{R}} 
\end{align}
where
\begin{align}
P_{mn\, \mathbf{R} \mathbf{d}} = \int_{S_\mathbf{d}} \mathbf{d}\cdot \mathbf{\mathcal{P}}_{mn}(\mathbf{r}) ~d s \label{eq:PRd}
\end{align}
\end{subequations}
and $\Delta_\mathbf{d}$ is the forward difference in the $\mathbf{d}$ direction defined by
$\Delta_\mathbf{d} f_{n\,\mathbf{R}} = \left[ f_{n\,{\mathbf{R}+\mathbf{d}}}- f_{n\,\mathbf{R}} \right] / |\mathbf{d}| $.
The discretized $k\cdot p$ term now consists of a link connecting sites multiplied by a coefficient defined on the link, which is therefore Hermitian. 
When approximated with naive finite differences, $\nabla \cdot a_{mn} (\mathbf{r}) \nabla f_n(\mathbf{r})$ also becomes non-Hermitian. 
Applying the same methods as above gives
\begin{subequations}
\begin{align}
\int_{\Omega_\mathbf{R}} \nabla \cdot a_{mn}(\mathbf{r}) \nabla f_n(\mathbf{r}) ~d^3r\approx
 \sum_\mathbf{d} A_{mn\, \mathbf{Rd}} \Delta_\mathbf{d} \, f_{n\,\mathbf{R}}\\
 A_{mn\, \mathbf{Rd}} = \int_{S_\mathbf{d}} a_{mn}(\mathbf{r}) ds \label{eq:ARd} .
\end{align}
\end{subequations}

Using the finite volume method, a derivative term becomes a sum of differences between sites multiplied by a coefficient that depends on an integral over the surface separating the cells centered on the two sites. Therefore the matrix element connecting two sites will be the same whether evaluated at $\mathbf{R}$ or $\mathbf{R}+\mathbf{d}$, making $H$ Hermitian.
The quantities given by Eq.s \ref{eq:ARd} and \ref{eq:PRd} do not need to be explicitly computed and we may
simply use the naive differencing formulas with coefficients empirically fit to bulk properties.
The finite volume method provides the justification for the coefficient being determined by the two sites being connected.

We may see explicitly how the finite volume method works by considering the Hamiltonian matrix elements involving the S and X states on two nearest neighbor sites at $\mathbf{R}_1$ and $\mathbf{R}_2$.
Using the differences given by Eq.s \ref{eq:dx1} and \ref{eq:dx2} in Eq. \ref{eq:HenvelopeDiscrete} we would have
\begin{equation}
H= \frac{1}{ a_{latt}} \frac{\hbar}{i m_0}\bordermatrix{
~&S_1 & X_1 & S_2 & X_2 \cr
~& 0 & 0 & 0 & \langle S| \hat p_x |X\rangle_{\Omega_1} \cr
~& 0 & 0 & \langle X| \hat p_x |S\rangle_{\Omega_1} & 0 \cr
~& 0 & -\langle S| \hat p_x |X\rangle_{\Omega_2} & 0 & 0 \cr
~& -\langle X| \hat p_x |S\rangle_{\Omega_2} & 0 & 0 & 0 \cr
}\label{eq:Hfvm1}
\end{equation}
which is clearly not Hermitian since all four of the non-zero matrix elements are different.
Using Eq. \ref{eq:HenvelopeContinuumAsym} will partially symmetrize Eq. \ref{eq:Hfvm1} because $\mathcal{P}_{mn} = -\mathcal{P}_{nm}$,
 but the matrix elements projected to $\Omega_1$ and $\Omega_2$ will still be different if the basis is inversion non-symmetric.
Applying the finite volume method then gives
\begin{equation}
H= \frac{1}{ a_{latt}} \bordermatrix{
~&S_1 & X_1 & S_2 & X_2 \cr
~& 0 & 0 & 0 & \mathcal{P}_{SX} \cr
~& 0 & 0 & \mathcal{P}_{XS} & 0 \cr
~& 0 & -\mathcal{P}_{SX} & 0 & 0 \cr
~& -\mathcal{P}_{XS} & 0 & 0 & 0 \cr
}\label{eq:Hfvm2}
\end{equation}
where we have omitted the coordinate indices on $\mathcal{P}$.
Since $\mathcal{P}_{mn}$ is real and $\mathcal{P}_{mn}=-\mathcal{P}_{nm}$, $H$ is Hermitian.

There are actually two separate symmetrizations:  Eq. \ref{eq:HenvelopeContinuumAsym} and the finite volume method. 
The former anti-symmetrizes the momentum matrix element with respect to the band index ("index symmtrization") and the latter symmetrizes with respect to the coordinate ("spatial symmetrization"). 
With an inversion symmetric Bloch basis only the index symmetrization is necessary to obtain a Hermitian Hamiltonian, but in the inversion non-symmetric case the finite volume method provides spatial symmetrization. 
The finite volume method results in equal momentum matrix elements, unlike the generalized Voronoi cell approach used in Sect. \ref{sect:four-bandModel} which gave two distinct S-X momentum matrix elements for an inversion non-symmetric basis.
It is interesting that Eq. \ref{eq:HenvelopeContinuumAsym} gives a diagonal kinetic term with the same form as the common operator ordering choice in the effective mass approximation, $\nabla \cdot \frac{1}{m(\mathbf{r})} \nabla \psi(\mathbf{r})$\cite{Zhu.prb.1983,Birman.book.1986,Einevoll.prb.1990}.

Note that if the Bloch functions centered on $\mathbf{R}$ and $\mathbf{R}+\mathbf{d}$ are different, there will be a discontinuity in the slope of $u_n(\mathbf{r})$ at the interface and the surface integrals in Eq.s \ref{eq:PRd} and \ref{eq:ARd} will be ill-defined, depending on whether we use $\mathbf{\mathcal{P}}_{mn}(\mathbf{r})$ and $a_{mn}(\mathbf{r})$ from the cell around $\mathbf{R}$ or $\mathbf{R}+\mathbf{d}$.
Since Bloch functions do not vary greatly among different III-V semiconductors,
we may take the surface integral to simply be a parameter depending on the atomic species at $\mathbf{R}$ and $\mathbf{R}+\mathbf{d}$.
Moreover, the true $u_m$ at a heterojunction will be a smooth function depending primarily on the type of the dimer, with some small dependence on the nearby atoms.
Taking the coefficient to depend only on the two atoms connected is essentially ignoring the influence of nearby atoms on the microscopic $V_0(\mathbf{r})$ and assuming the Bloch function at a dimer is the same as what it would be for that dimer in a bulk binary material.
The discontinuity in the Bloch functions at a heterojunction potentially spoils current conservation which requires $\psi(\mathbf{r})$ have continuous first derivatives. A discontinuity in the slope of $u_m$ requires a compensating discontinuity in the slope of the associated envelope.
Such an envelope is certainly possible in the continuum, however this cannot be accomplished in a real-space formulation on a grid since such a discontinuity would be over a length scale smaller than the cutoff imposed by the grid.
The Burt-Foreman formulation of envelope theory resolves this problem and ensures current conservation, but requires additional momentum matrix elements between wave vectors outside the first Brillouin zone\cite{Burt.sst.1988a,Foreman.prb.1996}.

\section{eight-band model}
\label{sect:eight-bandModel}
We now apply the ideas developed in Secs. \ref{sect:introduction}-\ref{sect:finiteVolumeMethod} to the eight-band Kane model with spin-orbit coupling and perturbative remote band contributions.
This model has been used to describe electronic states in bulk materials, impurities, and nanostructures.
The Hamiltonian is given by\cite{Pidgeon.pr.1966,Kane.jpcs.1957,Bahder.prb.1990}
\begin{align}
\label{eq:Bahder}
H_8 &= \nonumber \\
&{\let\quad\thinspace 
\bordermatrix{
~& u^{\Gamma_6}_{-1/2} & u^{\Gamma_6}_{1/2} &u^{\Gamma_8}_{+1/2} & u^{\Gamma_8}_{+3/2} & u^{\Gamma_8}_{-3/2} & u^{\Gamma_8}_{-1/2} &u^{\Gamma_7}_{-1/2} & u^{\Gamma_7}_{+1/2}\cr
~& A & 0 & T^*+V^* & 0 & -\sqrt{3}(T-V) & \sqrt{2}(W-U) & (W-U) & \sqrt{2}(T^*+V^*) \cr
~& 0 & A & \sqrt{2}(W-U) & -\sqrt{3}(T^*+V^*) & 0 & T-V & -\sqrt{2}(T-V) & W^*+U \cr
~& (T+V) & \sqrt{2}(W^*-U) & -P+Q & -S^* & R & 0 & \sqrt{\frac{3}{2}}S & -\sqrt{2}Q \cr
~& 0 & -\sqrt{3}(T+V) & -S & -P-Q & 0 & R & -\sqrt{2}R & \frac{1}{\sqrt{2}} S\cr
~& -\sqrt{3}(T^*-V^*) & 0 & R^* & 0 & -P-Q & S^* & \frac{1}{\sqrt{2}}S^* & \sqrt{2}R^* \cr
~& \sqrt{2}(W^*-U) & T^*-V^* & 0 & R^* & S & -P+Q & \sqrt{2}Q & \sqrt{\frac{3}{2}} S^* \cr
~& W^*-U & \sqrt{2}(T^*-V^*) & \sqrt{\frac{3}{2}}S^* & -\sqrt{2}R^* & \frac{1}{\sqrt{2}}S & \sqrt{2}Q & Z & 0 \cr
~& \sqrt{2}(T+V) & W+U & -\sqrt{2}Q & \frac{1}{\sqrt{2}}S^* & \sqrt{2}R & \sqrt{\frac{3}{2}} S & 0 & Z \cr
}}
\end{align}
where
\begin{align}
A&=E_c+(1+2 F)\frac{\hbar^2}{2 m_0} k^2,\nonumber \\
U&=\frac{1}{\sqrt{3}}P_0 k_z,\nonumber \\
V&=\frac{1}{\sqrt 6}P_0(k_x-i k_y),\nonumber \\
W&=\frac{i}{\sqrt 3} B k_x k_y,\nonumber \\
T&=\frac{1}{\sqrt 6}B k_z (k_x + i k_y ),\\
P&=-E_v + \frac{\hbar^2}{2m_0}\gamma_1 k^2,\nonumber \\
Q&=\frac{\hbar^2}{2m_0} \gamma_2 k^2,\nonumber \\
R&= -{\sqrt 3} \frac{\hbar^2}{2m_0} \left( \gamma_2 ( k_x^2- k_y^2 )- i 2 \gamma_3 k_x k_y\right),\nonumber \\
S&=\sqrt{3}\gamma_3 \frac{\hbar^2}{m_0} k_z (k_x-i k_y ),\nonumber \\
Z&=E_s - \frac{\hbar^2}{2m_0}\gamma_1 k^2 \nonumber 
\end{align}
where $ E_c$, $ E_v$, and $ E_s = E_v - \Delta$ are the zone-center energies of the conduction, valence, and split-off bands respectively, $ F$ is the remote band contribution to the conduction band effective mass, $iP_0=\frac{\hbar}{m_0}\langle S | P_x | X \rangle$, and $B$ is the inversion asymmetry parameter due to remote bands, which we take to be zero.
The modified Luttinger parameters are given by
\begin{align}
\gamma_1 = \gamma^L_1 - \frac{E_P}{3 E_g},\nonumber \\
\gamma_2 = \gamma^L_2 - \frac{E_P}{6 E_g},\\
\gamma_3 = \gamma^L_3 - \frac{E_P}{6 E_g}.\nonumber 
\end{align}
Note that we use the definitions from Ref. \onlinecite{Pidgeon.pr.1966} rather than Ref. \onlinecite{Bahder.prb.1990} since they give the standard relationships between hole masses and Luttinger parameters,
\begin{align}
\left( \frac{m_0}{m_{hh}^*}\right)^{[100]} = \gamma^L_1 - 2 \gamma^L_2\nonumber \\
\left( \frac{m_0}{m_{lh}^*}\right)^{[100]} = \gamma^L_1 + 2 \gamma^L_2\nonumber\\
\left( \frac{m_0}{m_{hh}^*}\right)^{[111]} = \gamma^L_1 - 2 \gamma^L_3\\
\left( \frac{m_0}{m_{lh}^*}\right)^{[111]} = \gamma^L_1 + 2 \gamma^L_3\nonumber.
\end{align}
For actual III-V materials the modified Luttinger parameters in Ref. \onlinecite{Bahder.prb.1990} give effective masses that differ from the relationships given above by about $10\%$.

To take the atomistic limit of the eight-band model we proceed as in Sect. \ref{sect:four-bandModel}, replacing k's with the appropriate difference operators on the crystal lattice.
As in Sect. \ref{sect:four-bandModel} we must include the additional atomistic momentum matrix element $i Q_a =\frac{\hbar}{m_0} \langle X | p_y | Y \rangle_{\Omega_1}=-\frac{\hbar}{m_0} \langle X | p_y | Y \rangle_{\Omega_2} $ where $\Omega_1$ ($\Omega_2$) is the volume around the anion (cation).
With this sign convention a positive $Q_a$ will give a long wavelength 16-band model with $Q>0$ in agreement with Ref. \onlinecite{Cardona.prb.1988}.
With the inclusion of spin and transforming to the total angular momentum basis in which Eq. \ref{eq:Bahder} is written, we obtain
\begin{align}
H_Q &=\left(
\begin{array}{cccccccc}
0 & 0 & 0 & 0 & 0 & 0 & 0 & 0 \\
0 & 0 & 0 & 0 & 0 & 0 & 0 & 0 \\
0 & 0 & 0 & -\frac{1}{\sqrt{3}}Q_a k_- & \frac{1}{\sqrt{3}} Q_a k_z & 0 & -\frac{1}{\sqrt{2}} Q_a k_+ & 0 \\
0 & 0 & \frac{1}{\sqrt{3}} Q_ak_+ & 0 & 0 & \frac{1}{\sqrt{3}}Q_a k_z & -\sqrt{\frac{2}{3}}Q_a k_z & -\frac{1}{\sqrt{6}} Q_a k_+ \\
0 & 0 & -\frac{1}{\sqrt{3}}Q_a k_z & 0 & 0 & \frac{1}{\sqrt{3}}Q_a(k_x-i k_y) & \frac{1}{\sqrt{6}} Q_ak_- & -\sqrt{\frac{2}{3}} Q_ak_z \\
0 & 0 & 0 & -\frac{1}{\sqrt{3}}Q_a k_z & -\frac{1}{\sqrt{3}}Q_a k_+ & 0 & 0 & \frac{1}{\sqrt{2}}Q_a k_- \\
0 & 0 & \frac{1}{\sqrt{2}}Q_a k_- & \sqrt{\frac{2}{3}}Q_a k_z & -\frac{1}{\sqrt{6}}Q_a k_+ & 0 & 0 & 0 \\
0 & 0 & 0 & \frac{1}{\sqrt{6}} Q_a k_- & \sqrt{\frac{2}{3}}Q_a k_z & -\frac{1}{\sqrt{2}}Q_a k_+ & 0 & 0 
\end{array}
\right)
\end{align}
where $k_\pm=(k_x\pm i k_y)$.
As in the case of the four-band model, $H_Q$ is the atomistic contribution on the anion site with $-H_Q$ on the cation site.

Rather than using modified Voronoi cells as we did with the four-band model, here we avoid the Hermiticity problem by choosing a basis of inversion symmetric Bloch functions. 
The eight-band model with $B=0$ is inversion symmetric, so adopting such a basis is quite natural, and in any case the choice of basis is arbitrary.
Although the basis is symmetric, inversion symmetry is broken by $\hat H_0$ (c.f. Eq. \ref{eq:H0}) which will cause the envelope functions to be inversion non-symmetric.
The eight-band model depends on eight parameters ($E_c$, $E_v$, $E_s$, $P_0$, $\gamma_1$, $\gamma_2$, $\gamma_3$, and $F$). 
As discussed in Sect. \ref{sect:four-bandModel}, the diagonal energies will be different on each atom, doubling their number to six. 
The other parameters, which depend on derivatives of the inversion symmetric Bloch functions, will be the same on each atom.
With the inclusion of $Q_a$, the number of parameters is increased from eight to 12 in the atomistic limit.

\begin{figure}
\includegraphics[width=0.5\columnwidth]{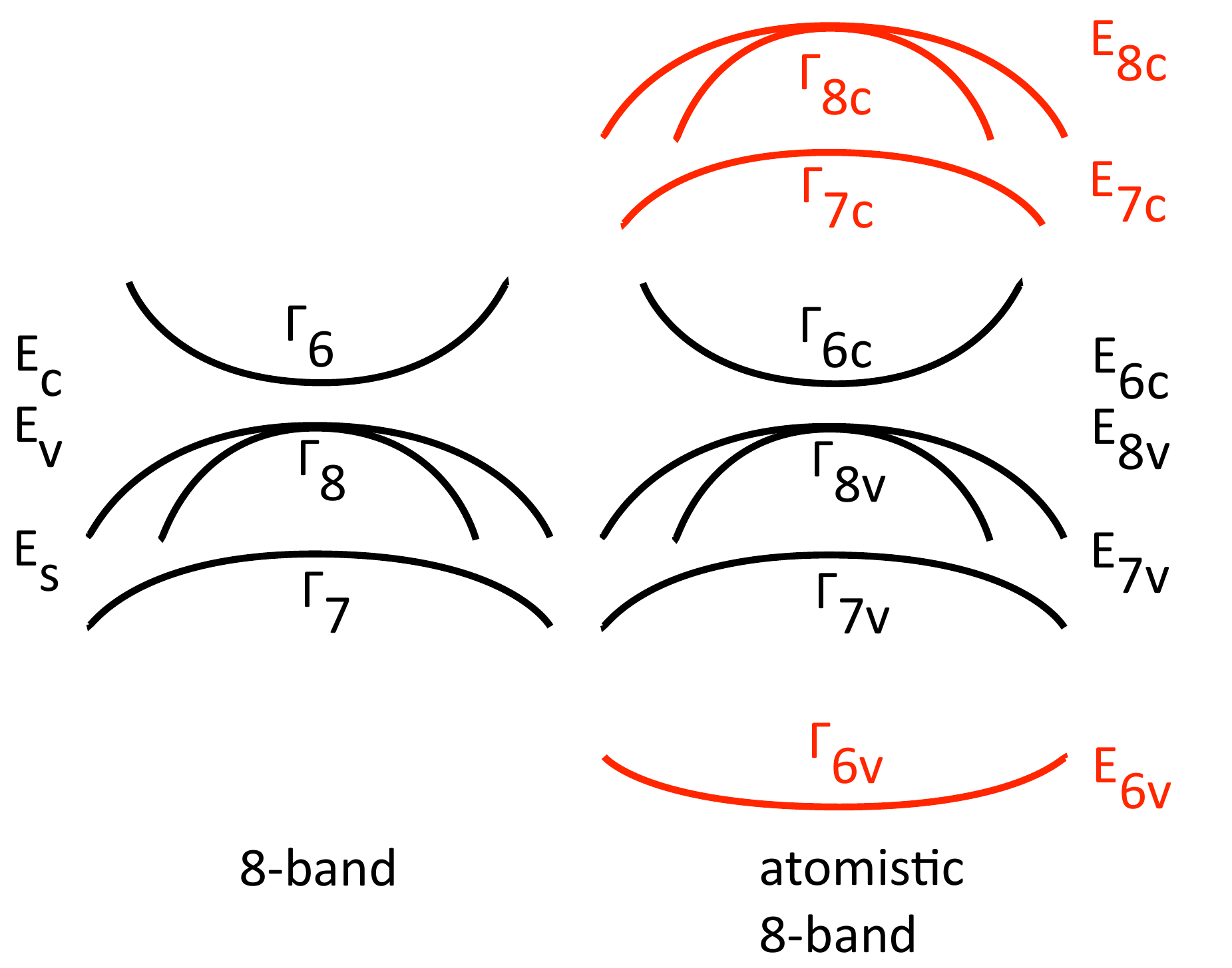}
\caption{In the atomistic limit the number of bands is doubled due to there being two atoms per unit cell. 
For each band of the original continuum $k\cdot p$ model there will be one state with an envelope that has the same sign on both atoms ($++$) and another with an envelope having opposites signs ($+-$) on the two atoms.
The $++$ solutions are taken to be the states of the original continuum $k\cdot p$ model and the $+-$ solutions, which are shifted in energy by $\approx \mathcal{O}(\hbar^2/m_0a_{latt}^2)$, are taken to be excited bands.
The energy of an excited band also depends on the remote band contribution to the effective mass (i.e. the coefficient of $\nabla^2$ in the atomistic Hamiltonian).
The remote contribution to the valence band is multiplied by $\gamma_{a1}>0$, shifting the $\Gamma_{8c}$ and $\Gamma_{7c}$ states to higher energies.
Because the remote contribution to the conduction band mass is negative the $\Gamma_{6v}$ band is displaced downward.}
\label{fig:bandSchematic}
\end{figure}

The remote band contributions in the eight-band model introduce additional $k$s, and thus additional difference operators.
As discussed in Sect. \ref{sect:atomisticGrid}, the fact that each atom has 4 neighbors means that only $k^2$, $k_x$, $k_y$, and $k_z$ can be constructed using nearest neighbor differences while $k_x k_y$, $k_x k_z$, $k_y k_z$, $k_x^2$, $k_y^2$, and $k_z^2$ require next nearest neighbor differences.
Replacing the $k$s with difference operators acting on plane waves as in Sect. \ref{sect:four-bandModel}, the Hamiltonian becomes a $16\times 16$ matrix of the form
\begin{align}
\left(
\begin{array}{cc}
 H_{11}& H_{12} \\
 H_{12}^\dagger & H_{22}\\
\end{array}
\right)
\end{align}
where the diagonal blocks $H_{11}$ and $H_{22}$ include on-site and next-nearest-neighbor couplings and the $H_{12}$ block contains nearest neighbor couplings.
For $k=0$ the nearest neighbor couplings from $k_x$, $k_y$, and $k_z$ vanish but those from $k^2$ remain, shifting the zone-center energies from their eight-band values.
This can be seen in Eq. \ref{eq:atomistickp4} where $\mathcal T$s are nonzero even for $\mathbf k = 0$.
For zincblende the onsite energies are different on the two atoms ($E_{c1}$, $E_{v1}$, $E_{s1}$ on the anion and $E_{c2}$, $E_{v2}$, $E_{s2}$ on the cation),
breaking the inversion symmetry.
With two grid sites per unit cell the number of bands in the original $k\cdot p$ model is doubled in the atomistic limit and each zone-center state of the continuum model splits into two states: one having an envelope with the same sign on each atom ($++$) and another with opposite signs ($+-$).
The $++$ eigenvectors correspond to the original eight-band states ($\Gamma_{6c}$, $\Gamma_{7v}$, $\Gamma_{8v}$) and the $+-$ states to correspond to the additional states found in the 16-band model ($\Gamma_{6v}$, $\Gamma_{8c}$, $\Gamma_{7c}$) as shown in Fig. \ref{fig:bandSchematic}.

To understand the band doubling, consider a continuum Hamiltonian having a diagonal term $H=E_c + C_c k^2$ where the subscripts indicate the parameters are for the continuum model.
The parameters of the atomistic model are determined by fitting to the parameters of the continuum model.
Because $\nabla^2$ couples the two atoms within a unit cell the atomistic Hamiltonian is a non-diagonal $2\times 2$ matrix even at $\mathbf{k}=0$,
\begin{align}
\left(
\begin{array}{cc}
 E_{a1} + \frac{32}{a_{latt}^2} C_a & -\frac{32}{a_{latt}^2} C_a \\[8pt]
 - \frac{32}{a_{latt}^2} C_a & E_{a2} + \frac{32}{a_{latt}^2} C_a \\
\end{array}
\right)
\end{align}
where $E_{a1}$, $E_{a2}$ are the atomistic energies for atom $1$ and $2$ (anion and cation respectively for zincblende) and $C_a$ is the atomistic coefficient corresponding to $C_c$ in the continuum Hamiltonian.
The zone-center eigenvalues and eigenvectors are then
\begin{subequations}
\begin{align}
E&= \frac{32}{a_{latt}^2}C_a +\frac{E_{a1}+E_{a2}}{2}\pm \frac{1}{2}\sqrt{(E_{a2}-E_{a1})^2+4096 C_a^2/a_{latt}^4} \label{eq:atomisticZoneCenterE}\\
f &=
\left(
\begin{array}{c}
 A / \sqrt{1+A^2}\\
 1/\sqrt{1+A^2}
\end{array}
\right)\label{eq:atomisticEigenvector}\\
A&= \frac{a_{latt}^2}{64 C_a} \left( E_{a2}-E_{a1} \right)\mp \sqrt{\frac{a_{latt}^4}{4096 C_a^2} (E_{a2}-E_{a1})^2 +1} .
\end{align}
\end{subequations}

In the inversion symmetric case, $E_{a1}=E_{a2}$ and the zone center energies are given by $E=E_a$ and $E=E_a+32 C_a/a_{latt}^2$.
The atomistic parameter $E_a$ is then set equal to the continuum zone center energy, $E_c$, and the atomistic limit gives rise to an additional band with zone-center energy $32 C_a/a_{latt}^2$ higher.
If $C_a$ is fixed by fitting to an effective mass, there are no additional parameters to fit.
In the inversion non-symmetric case the zone center energies of the atomistic model are functions of three parameters, $E_{a1}$, $E_{a2}$, and $C_a$, the last of which would be fixed by fitting to an effective mass.
We require two conditions to fit $E_{a1}$ and $E_{a2}$.
The simplest approach is to set Eq. \ref{eq:atomisticZoneCenterE} to $E_c$ and the energy of one additional band.
Alternatively, one could fit to $E_c$, and another condition such as the ratio of of envelope functions on the two atoms.
Since empirical data for energies of excited bands is more readily available than Bloch functions, we will determine $E_{a1}$, and $E_{a2}$ by fitting to the two band energies.
Our fitting procedure is by no means unique, and additional data may make some other method preferable.

We take as our target energies those of the 16-band model, $E_{6v}$, $E_{7v}$, $E_{8v}$, $E_{6c}$, $E_{7c}$, $E_{8c}$ (see Fig. \ref{fig:bandSchematic}). These have been measured for some materials \cite{Aspnes.prb.1973} and calculated for others\cite{Chelikowsky.prb.1976}.
Setting the zone-center eigenvalues of the atomistic Hamiltonian equal to the target energies gives the atomistic on-site energies
\begin{align}
E_{c1}&=\frac{1}{2}(E_{6c}+E_{6v}) - 16 (1+2 F_a) \hbar^2/m_0 a_{latt}^2- \frac{1}{2} \sqrt{(E_{6c}-E_{6v})^2 -1024 (1+2 F_a)^2 \hbar^4/m_0^2 a_{latt}^4} \nonumber \\
E_{c2}&=\frac{1}{2}(E_{6c}+E_{6v}) - 16 (1+2 F_a) \hbar^2/m_0 a_{latt}^2+ \frac{1}{2} \sqrt{(E_{6c}-E_{6v})^2 -1024 (1+2 F_a)^2 \hbar^4/m_0^2a_{latt}^4} \nonumber \\
E_{s1}&=\frac{1}{2}(E_{7c}+E_{7v}) + 16 \gamma_{a1} \hbar^2/m_0 a_{latt}^2- \frac{1}{2} \sqrt{(E_{7c}-E_{7v})^2 -1024 \gamma_{a1}^2 \hbar^4/m_0^2a_{latt}^4} \nonumber \\
E_{s2}&=\frac{1}{2}(E_{7c}+E_{7v}) + 16 \gamma_{a1} \hbar^2/m_0 a_{latt}^2+ \frac{1}{2} \sqrt{(E_{7c}-E_{7v})^2 -1024 \gamma_{a1}^2 \hbar^4/m_0^2a_{latt}^4} \nonumber \\
E_{v1}&=\frac{1}{2}(E_{8c}+E_{8v}) + 16 \gamma_{a1} \hbar^2/m_0 a_{latt}^2- \frac{1}{2} \sqrt{(E_{8c}-E_{8v})^2 -1024 \gamma_{a1}^2 \hbar^4/m_0^2a_{latt}^4} \nonumber \\
E_{v2}&=\frac{1}{2}(E_{8c}+E_{8v}) + 16 \gamma_{a1} \hbar^2/m_0 a_{latt}^2+ \frac{1}{2} \sqrt{(E_{8c}-E_{8v})^2 -1024 \gamma_{a1}^2 \hbar^4/m_0^2a_{latt}^4} \label{eq:As}
\end{align}
where $E_{c1}$ is the on-site conduction band energy on the atom at the origin (anion for zincblende), $E_{c2}$ is the on-site energy for the atom at $(a_{latt}/4,a_{latt}/4,a_{latt}/4)$ (cation for zincblende), and likewise for the valence band (subscripts $v1$ and $v2$) and the spin-orbit band (subscripts $s1$ and $s2$).
The $a$ subscripts on $F_a$ and $\gamma_{a1}$ indicate they are the atomistic versions of the continuum $k\cdot p$ parameters $F$ and $\gamma_1$.
The corresponding envelope functions are obtained from Eq. \ref{eq:atomisticEigenvector}, giving
\begin{align}
A_{6v} &= \left((E_{c2}-E_{c1}) m_0 a_{latt}^2 + \sqrt{1024 (1+2 F_a)^2 \hbar^4 +(E_{c2}-E_{c1})^2 m_0^2 a_{latt}^4} \right)/{32 (1+2 F_a) \hbar^2} \nonumber \\
A_{6c} &= \left((E_{c2}-E_{c1}) m_0 a_{latt}^2 - \sqrt{1024 (1+2 F_a)^2 \hbar^4 +(E_{c2}-E_{c1})^2 m_0^2 a_{latt}^4} \right)/{32 (1+2 F_a) \hbar^2} \nonumber \\
A_{7v} &= -\left((E_{s2}-E_{s1}) m_0 a_{latt}^2 + \sqrt{1024 \gamma_{a1}^2 \hbar^4 +(E_{s2}-E_{s1})^2 m_0^2 a_{latt}^4} \right)/{32 \gamma_{a1} \hbar^2} \nonumber \\
A_{7c} &= -\left((E_{s2}-E_{s1}) m_0 a_{latt}^2 - \sqrt{1024 \gamma_{a1}^2 \hbar^4 +(E_{s2}-E_{s1})^2 m_0^2 a_{latt}^4} \right)/{32 \gamma_{a1} \hbar^2} \nonumber \\
A_{8v} &= -\left( (E_{v2}-E_{v1}) m_0 a_{latt}^2 + \sqrt{1024 \gamma_{a1}^2 \hbar^4 +(E_{v2}-E_{v1})^2 m_0^2 a_{latt}^4}\right)/{32 \gamma_{a1} \hbar^2} \nonumber \\
A_{8c} &= -\left( (E_{v2}-E_{v1}) m_0 a_{latt}^2 - \sqrt{1024 \gamma_{a1}^2 \hbar^4 +(E_{v2}-E_{v1})^2 m_0^2 a_{latt}^4} \right)/{32 \gamma_{a1} \hbar^2}.
\end{align}
The states with energies $E_{6c}$, $E_{7v}$, $E_{8v}$ have envelopes of the form $++$ while those with energies $E_{6v}$, $E_{8c}$, $E_{7c}$ have $+-$ envelopes.
Note that the envelopes break inversion symmetry and the momentum matrix elements taken over a unit cell must include the sub-unit cell structure of the envelopes.

Having determined the on-site energies, we must now determine the coefficients of the difference operators.
In $k\cdot p$ theory effective masses only require knowing $E(k)$ to second order in $k$, and are therefore computed using second order perturbation theory in $k$\cite{Hermann.prb.1977,Pfeffer.prb.1996,Roth.pr.1959}.
The g-factor is also computed this way\cite{Roth.pr.1959}.
For the 16-band continuum $k\cdot p$ model the results are
\begin{subequations}
\begin{align}
 \frac{m_0}{m^*}&=(1+2 F)+\frac{E_{P_0}}{3} \left( \frac{2}{E_{6c}-E_{8v}} +\frac{1}{E_{6c}-E_{7v}} \right)
			 +\frac{E_{P_1}}{3} \left( \frac{2}{E_{6c}-E_{8c}} +\frac{1}{E_{6c}-E_{7c}} \right) \label{eq:contmassA} \\ 
\frac{g^*}{g_0} &= 1 + \frac{g_r}{g_0} - \frac{E_{P_0}}{3}\left( \frac{1}{E_{6c}-E_{8v}}-\frac{1}{E_{6c}-E_{7v}} \right)
		 - \frac{E_{P_1}}{3}\left( \frac{1}{E_{6c}-E_{8c}}-\frac{1}{E_{6c}-E_{7c}} \right) \label{eq:contmassB} \\
\gamma_1^L &= \gamma_1 - \frac{E_{P_0}}{3 (E_{8v}-E_{6c})} - \frac{E_{P_2}}{3 (E_{8v}-E_{6v})} - \frac{E_Q}{3(E_{8v}-E_{7c})} - \frac{E_Q}{3(E_{8v}-E_{8c})} \label{eq:contmassC} \\
\gamma_2^L &= \gamma_2 - \frac{E_{P_0}}{6 (E_{8v}-E_{6c})} - \frac{E_{P_2}}{6 (E_{8v}-E_{6v})} + \frac{E_Q}{6(E_{8v}-E_{7c})}\label{eq:contmassD} \\
\gamma_3^L &= \gamma_3 - \frac{E_{P_0}}{6 (E_{8v}-E_{6c})} - \frac{E_{P_2}}{6 (E_{8v}-E_{6v})} - \frac{E_Q}{6(E_{8v}-E_{7c})}\label{eq:contmassE} 
\end{align}
\end{subequations}
where $g_0$ is the bare electron g factor, $g_r$ is a possible remote band contribution, $E_{P_0} =2| \langle S_c | P_x | X_v \rangle|^2/m_0$, $E_{P_1} = 2| \langle S_c | P_x | X_c \rangle|^2/m_0$, $E_{P_2} = 2| \langle S_v | P_x | X_v \rangle|^2/m_0$, and $E_{Q} = 2| \langle X_c | P_y | Z_v \rangle|^2/m_0$.
The quantities on the left hand sides of Eq.s \ref{eq:contmassA}-\ref{eq:contmassE} are the target parameters taken from experiment or possibly {\it ab initio} calculations while $ F$, $E_{P_0}$, $E_{P_1}$, $E_Q$, $\gamma_1$, $\gamma_2$, and $\gamma_3$ are the model parameters empirically chosen to reproduce the target values.
The model parameters are easily determined since the equations are linear. 
Since $E_{8v}-E_{6v}$ is large, we may set $E_{P_2}=0$, giving five equations in five unknowns.

The effective masses and g-factors of the atomistic model may also be computed using perturbation theory for small $k$.
The resulting expressions contain effective matrix elements that depend on the variation of the envelope over the unit cell and the bands which they connect.
For example, the momentum matrix element between the $\Gamma_{6c}$ and $\Gamma_{8v}$ states is
\begin{align}
P_{6c\,8v}&= -i \frac{\hbar}{m_0}\langle S_{\Gamma_{6c}}| P_x |X_{\Gamma_{8v}}\rangle\nonumber \\
&= \frac{1}{\sqrt{1+A_{6c}^2}}\left(
\begin{array}{c}
		A_{6c}\\
	 1
\end{array}
\right)^T
\left(
\begin{array}{cc}
0 & iP_a \\
iP_a &0 \\
\end{array}
\right)
 \frac{1}{\sqrt{1+A_{6c}^2}} \left(
\begin{array}{c}
		A_{8v}\\
	 1
\end{array}
\right) \nonumber \\
& = P_a (A_{6c}+A_{8v}) / \sqrt{(1+A_{6c}^2)(1+A_{8v}^2)}
\end{align}
with the other matrix elements ($P_{6c\,8v}$, $P_{6c\,7v}$, $P_{6c\,8c}$, $P_{6c\,7c}$, $P_{6v\,8v}$, $P_{6v\,7v}$, $P_{6v\,8v}$, and $P_{6v\,7v}$) defined similarly. The $Q$ matrix elements take the form
\begin{align}
Q_{8c\,8v}&= -i \frac{\hbar}{m_0}\langle X_{\Gamma_{8c}} | P_y | Z_{\Gamma_{8v}} \rangle = \frac{(A_{8c}-A_{8v}) Q_a}{\sqrt{(1+A_{8c}^2)(1+A_{8v}^2)}}.
\end{align}
Doing second order perturbation theory directly on the atomistic Hamiltonian we obtain
\begin{subequations}
\begin{align}
\frac{m_0}{m_c^*} &= \frac{2 A_{6c}(1+2 F_a)}{1+A_{6c}^2}
 + \frac{1}{3}\left( \frac{2E_{P\,6c\,8v}}{E_{6c}-E_{8v}} + \frac{E_{P\,6c\,7v}}{E_{6c}-E_{7v}} \right)
 + \frac{1}{3}\left( \frac{2E_{P\,6c\,8c}}{E_{6c}-E_{8c}} + \frac{E_{P\,6c\,7c}}{E_{6c}-E_{7c}} \right) \label{eq:pertmassesA}\\
\frac{g^*}{g_0} &= 1+\frac{g_r}{g_0}
 - \frac{1}{3}\left( \frac{E_{P\,6c\,8v}}{E_{6c}-E_{8v}} - \frac{E_{P\,6c\,7v}}{E_{6c}-E_{7v}} \right)
 - \frac{1}{3}\left( \frac{E_{P\,6c\,8c}}{E_{6c}-E_{8c}} - \frac{E_{P\,6c\,7c}}{E_{6c}-E_{7c}} \right) \label{eq:pertmassesB}\\
\gamma_1^L &= \frac{2 A_{8v}}{1+A_{8v}^2}\gamma_{a1} 
	- \frac{E_{P\,8v\,6c}}{3(E_{8v}-E_{6c})} - \frac{E_{P\,8v\,6v}}{3(E_{8v}-E_{6v})} 
	- \frac{E_{Q\,8v\,7c}}{3(E_{8v}-E_{7c})} - \frac{E_{Q\,8v\,8c}}{3(E_{8v}-E_{8c})} - \frac{E_{Q\,8v\,7v}}{3(E_{8v}-E_{7v})} \label{eq:pertmassesC}\\
\gamma_2^L &= \gamma_{a2} - \frac{E_{P\,8v\,6c}}{6(E_{8v}-E_{6c})} - \frac{E_{P\,8v\,6v}}{6(E_{8v}-E_{6v})} +\frac{E_{Q\,8v\,7c}}{6(E_{8v}-E_{7c})} +\frac{E_{Q\,8v\,7v}}{6(E_{8v}-E_{7v})} \label{eq:pertmassesD} \\
\gamma_3^L &= \gamma_{a3} + \frac{A_{8v} Q_a a_{latt}}{6(1+A_{8v}^2)}
	- \frac{E_{P\,8v\,6c}}{6(E_{8v}-E_{6c})} - \frac{E_{P\,8v\,6v}}{3(E_{8v}-E_{6v})} - \frac{E_{Q\,8v\,7c}}{6(E_{8v}-E_{7c})} - \frac{E_{Q\,8v\,7v}}{6(E_{8v}-E_{7v})}\label{eq:pertmassesE}
\end{align}
\end{subequations}
where the band-specific Kane energies are $E_{Pmn}= 2 |\langle S_m | \hat p_x | X_n \rangle |^2 /m_0 =2 m_0 P_{mn}^2 / \hbar^2$ and $E_{Qmn} = 2 |\langle X_m | \hat p_y |Z_n \rangle |^2 /m_0= 2m_0 Q_{mn}^2 / \hbar^2$.
The perturbative expression are the same as in continuum $k\cdot p$ except that matrix elements are replaced with the effective matrix elements including variation of the envelope within the unit cell.
Unlike the continuum case, the dependence on $F_a$ and $\gamma_{a1}$ is nonlinear.

\section{Parameter Fitting}
\label{sect:parameterFitting}
To determine the atomistic parameters we adopt a set of target material parameters to which we fit the atomistic parameters (see Table I).
These values are known with varying degrees of certainty, with some taken from high precision measurements while others are obtained theoretically.
The basic eight-band parameters are taken from Ref. \onlinecite{Vurgaftman.jap.2001}.
The zone-center energies of the higher lying bands ($\Gamma_{6v}$, $\Gamma_{7c}$, $\Gamma_{8c}$) are not as well known and we have taken their values from the $k\cdot p$ calculation of Ref. \onlinecite{Jancu.prb.2005}, with the exception of GaAs for which we used the experimental values from Ref. \onlinecite{Aspnes.prb.1973}.
The values of the conduction band effective g factor, $g^*$, and the Dresselhaus spin splitting, $\gamma_c$, were also taken from Ref. \onlinecite{Jancu.prb.2005}.
Some modifications have been made for InAs.
The spin orbit coupling has been increased to $\Delta = 0.45~ \rm eV$ in order to be able to obtain $g^* =-14.9$ without having $\gamma_c > 100~  \rm eV \, \AA^3$ .
Alternatively, the $\Delta$ could be left unchanged and $g^*$ fit using a remote band contribution $g_r$.
The value of $\gamma_2^L$ for InAs has been reduced from $8.2$ to $7.5$ in order to avoid bands that cross the gap at large $k$, which is much less of a liberty than it may seem since the Luttinger parameters for InAs are poorly known\cite{Vurgaftman.jap.2001}.

\def\mymathhyphen{{\hbox{-}}}

The atomistic on-site potentials ($E_{c1}$, $E_{c2}$, $E_{v1}$, etc. in Table \ref{tab:akpparams}) are determined from Eq. \ref{eq:As} using the lattice constants and zone-center energies from Table \ref{tab:target-params}.
To fit the effective masses we must determine the values of $F_a$, $P_a$, $Q_a$, $\gamma_{a1}$, $\gamma_{a2}$, $\gamma_{a3}$ for which Eq. \ref{eq:pertmassesA}-\ref{eq:pertmassesE} match the empirical target values.
This is more difficult than the fitting procedure for a continuum $k\cdot p$ model because the effective momentum matrix elements and remote band contributions depend on $F_a$ and $\gamma_{a1}$.
We do a nonlinear fit on $F_a$ and $\gamma_{a1}$.
For particular values of $F_a$ and $\gamma_{a1}$, $P_a$ is determined by Eq. \ref{eq:pertmassesA}, $Q_a$ is determined by Eq. \ref{eq:pertmassesC}, and $\gamma_{a2}$ and $\gamma_{a3}$ are determined by Eq.s \ref{eq:pertmassesD} and Eq. \ref{eq:pertmassesE}.
 $F_a$ is then adjusted to make the resulting $g^*$ match the target value.
 This results in a curve in the $F_a \mymathhyphen \gamma_{a1}$ plane from which we pick the point at which the Dresselhaus spin splitting fits the target as well.
 We determine $\gamma_c$ by numerically computing the spin-splitting $E({\mathbf k})$ in the $110$ direction.
The range of $F_a$ and $\gamma_{a1}$ which must be numerically searched is reduced by the condition that the amplitudes $A_{6v},...,A_{8c}$ must be real and the solutions corresponding to $E(\Gamma_{6c})$, $E(\Gamma_{7v})$, $E(\Gamma_{8v})$ must have signature $++$. 
These conditions restrict the values to $-a_{latt}^2(E_{6c}-E_{6v})/32 < \left( 2 F_a+1 \right) < 0$ and $-a_{latt}^2(E_{8c}-E_{8v})/32 < \gamma_{a1} < 0$.

\begin{figure}
\includegraphics[width=0.3\columnwidth]{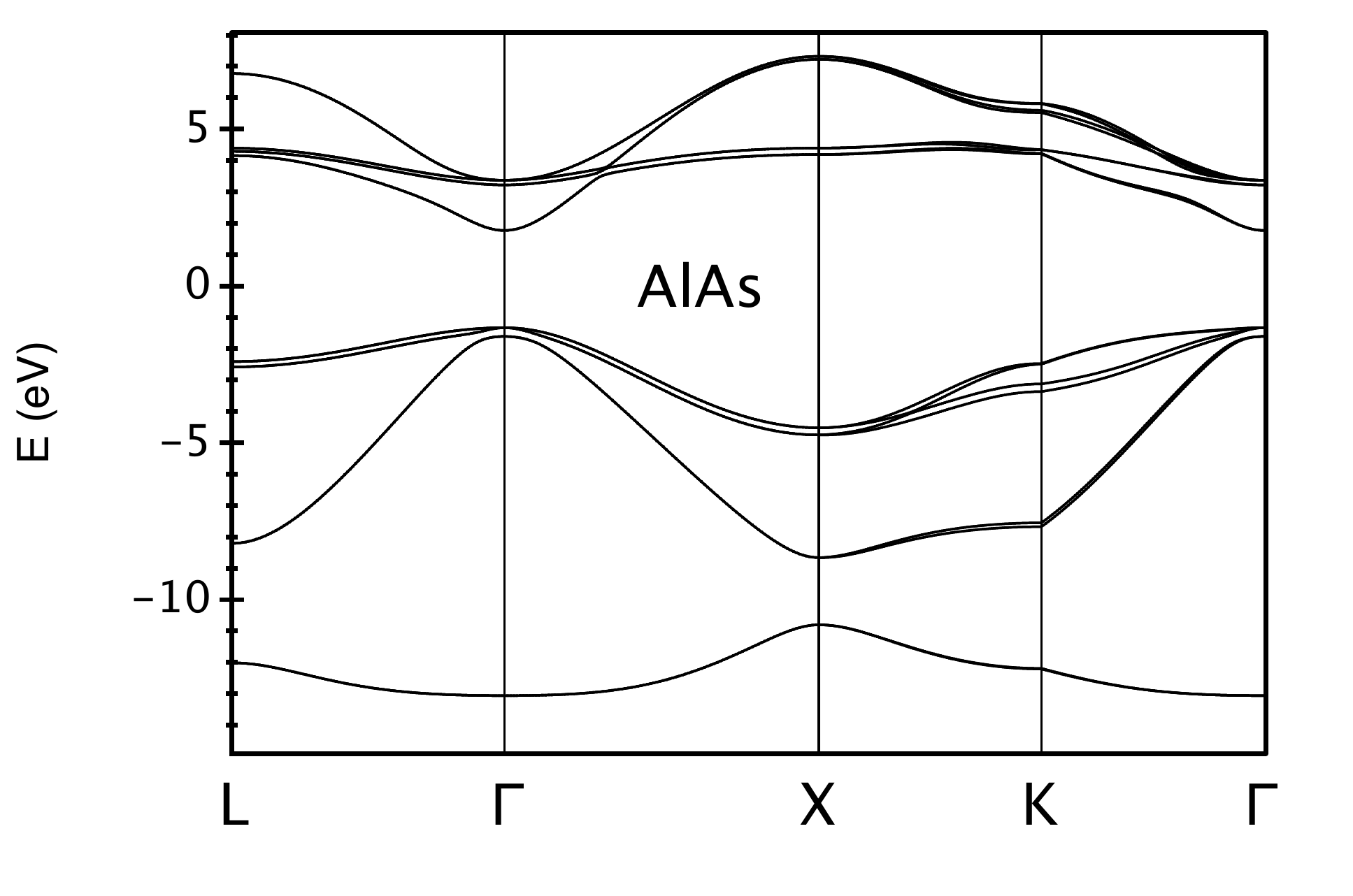}
\includegraphics[width=0.3\columnwidth]{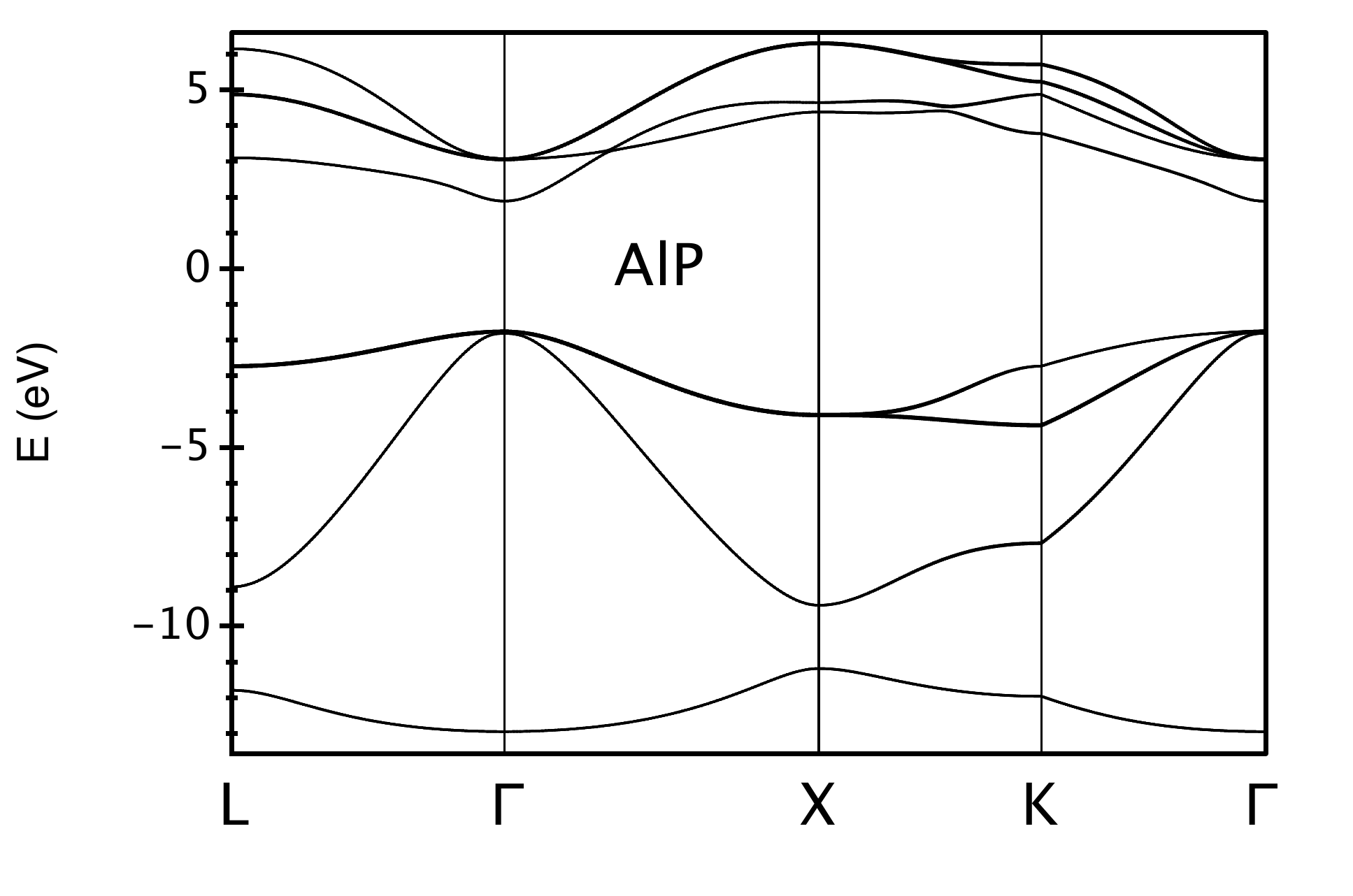}
\includegraphics[width=0.3\columnwidth]{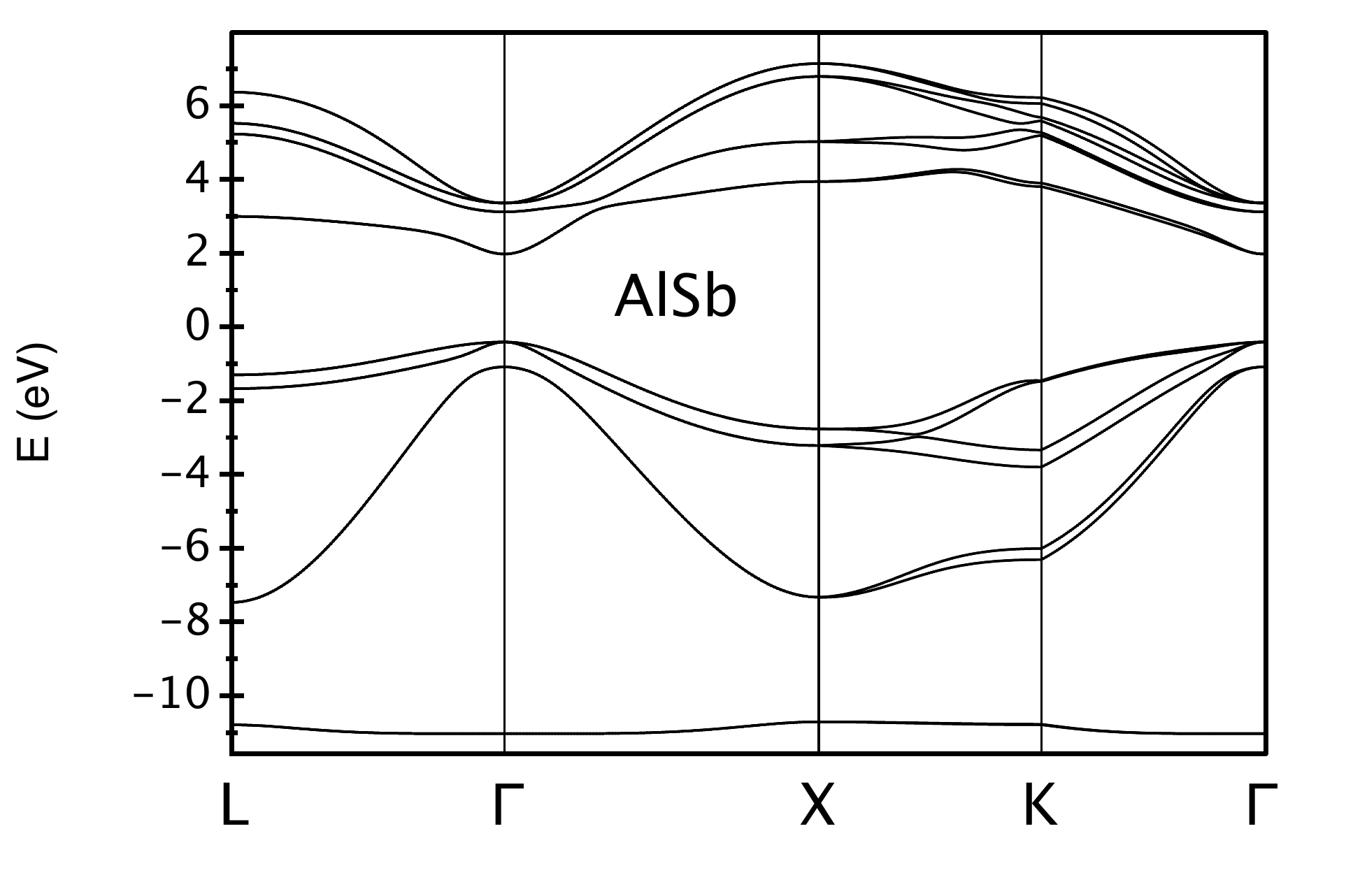}
\includegraphics[width=0.3\columnwidth]{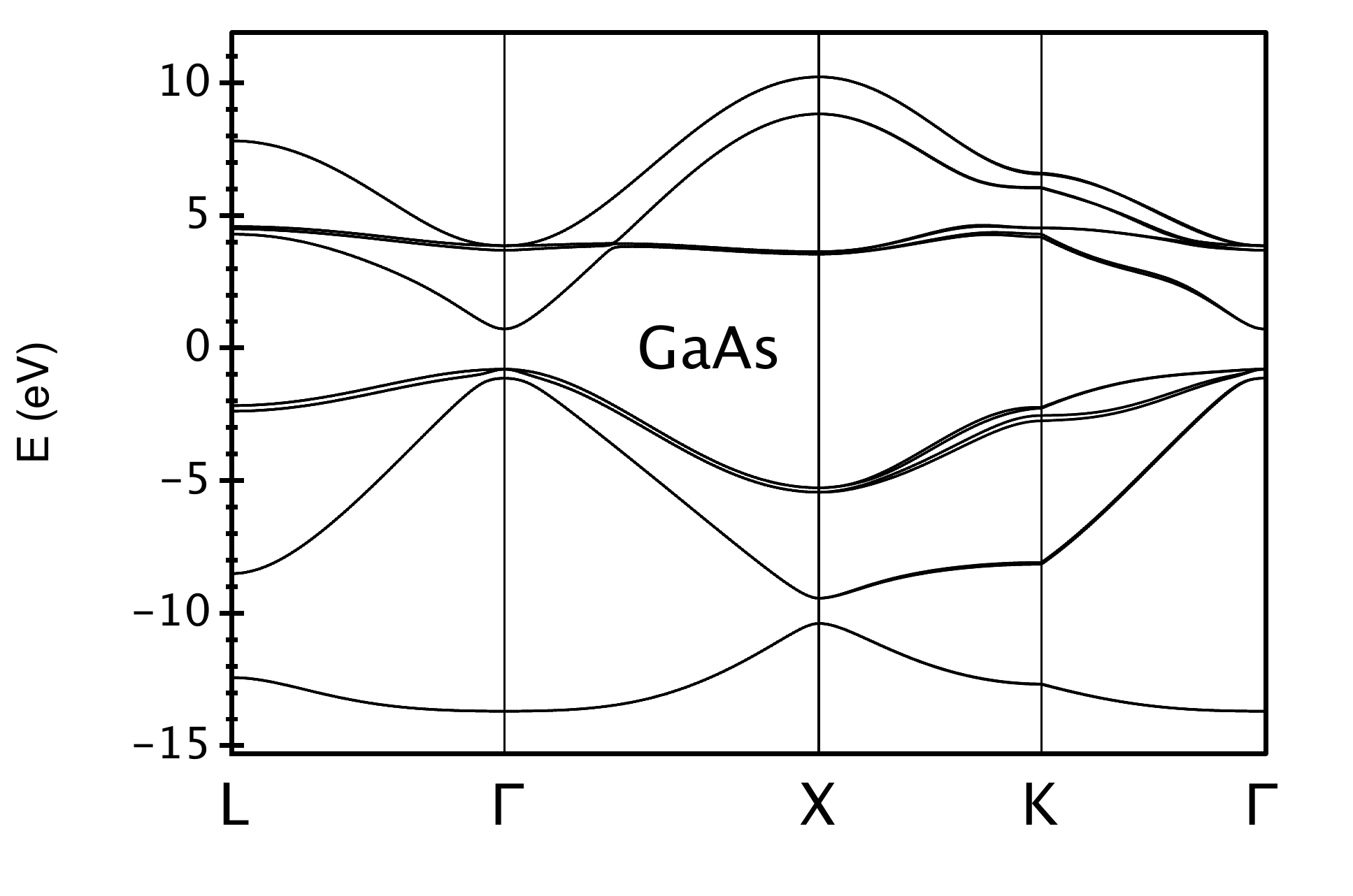}
\includegraphics[width=0.3\columnwidth]{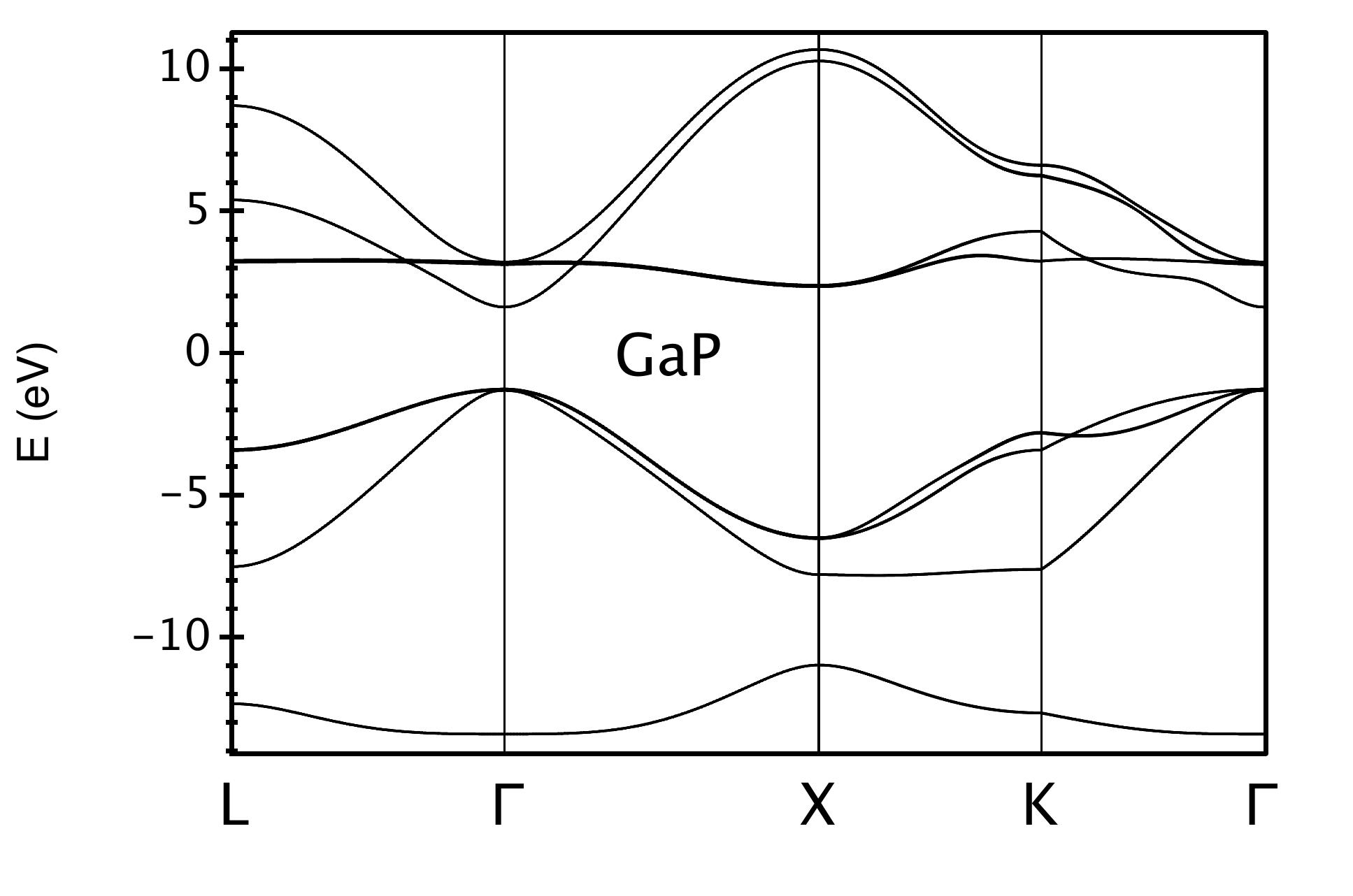}
\includegraphics[width=0.3\columnwidth]{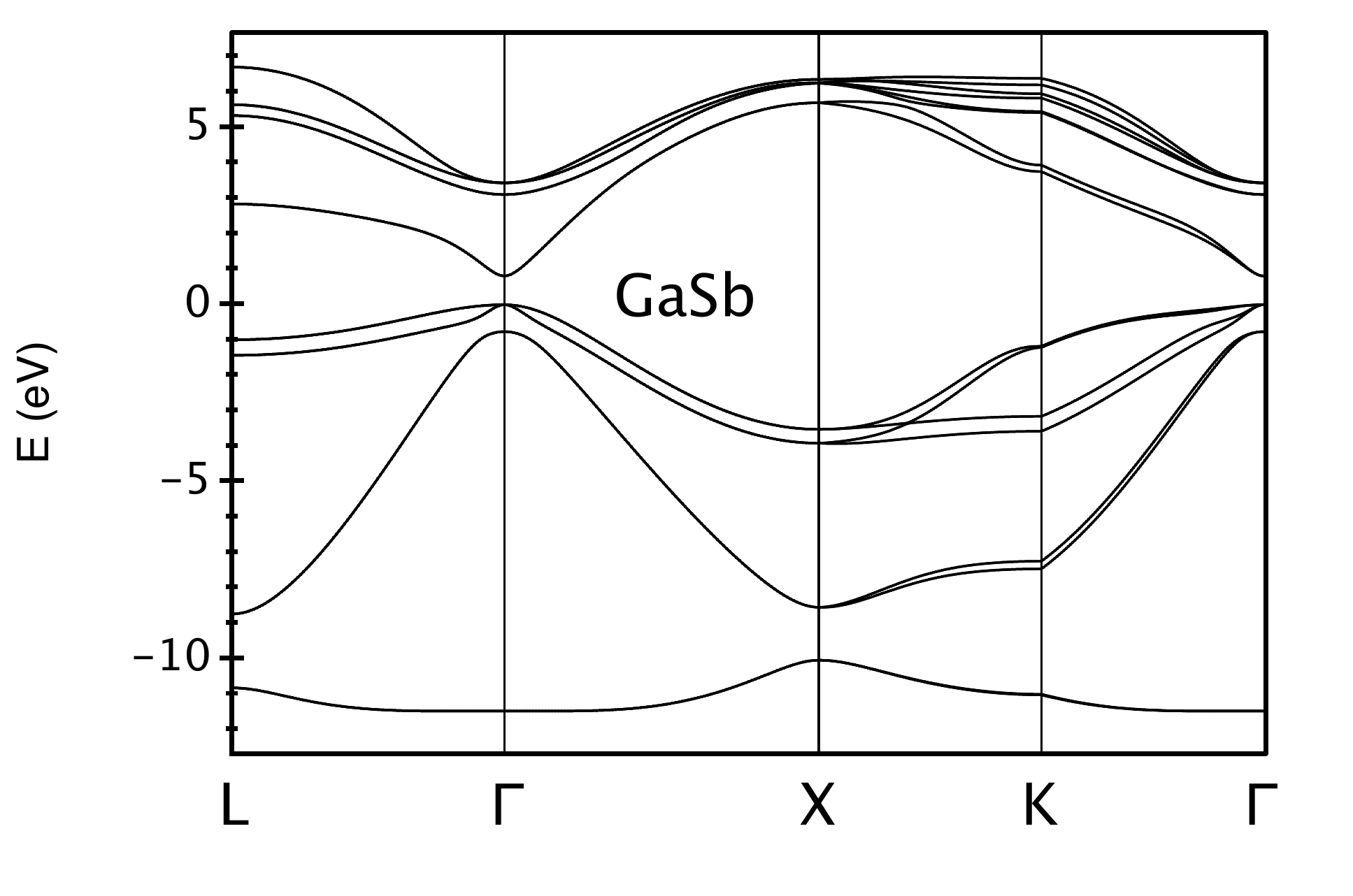}
\includegraphics[width=0.3\columnwidth]{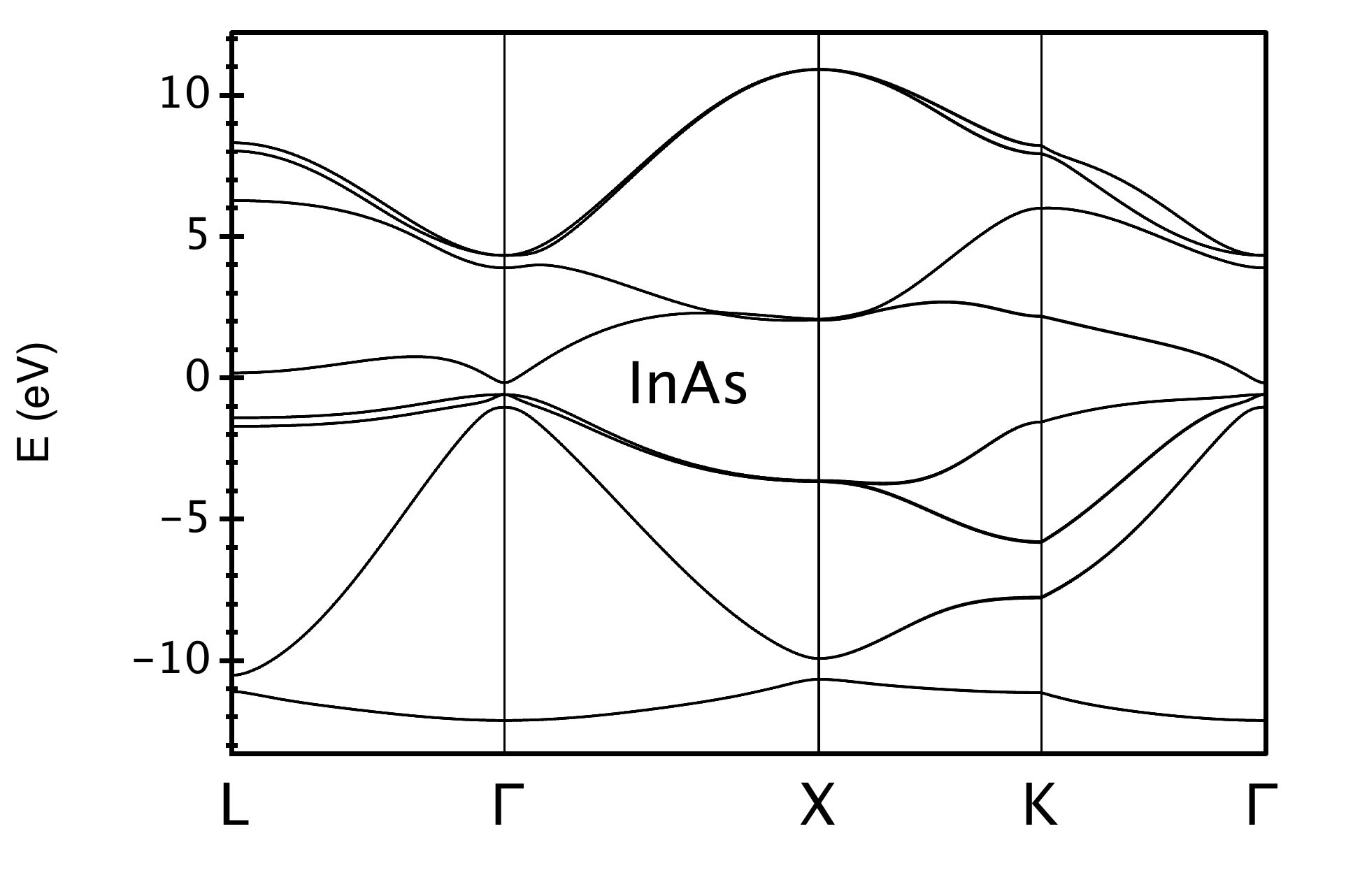}
\includegraphics[width=0.3\columnwidth]{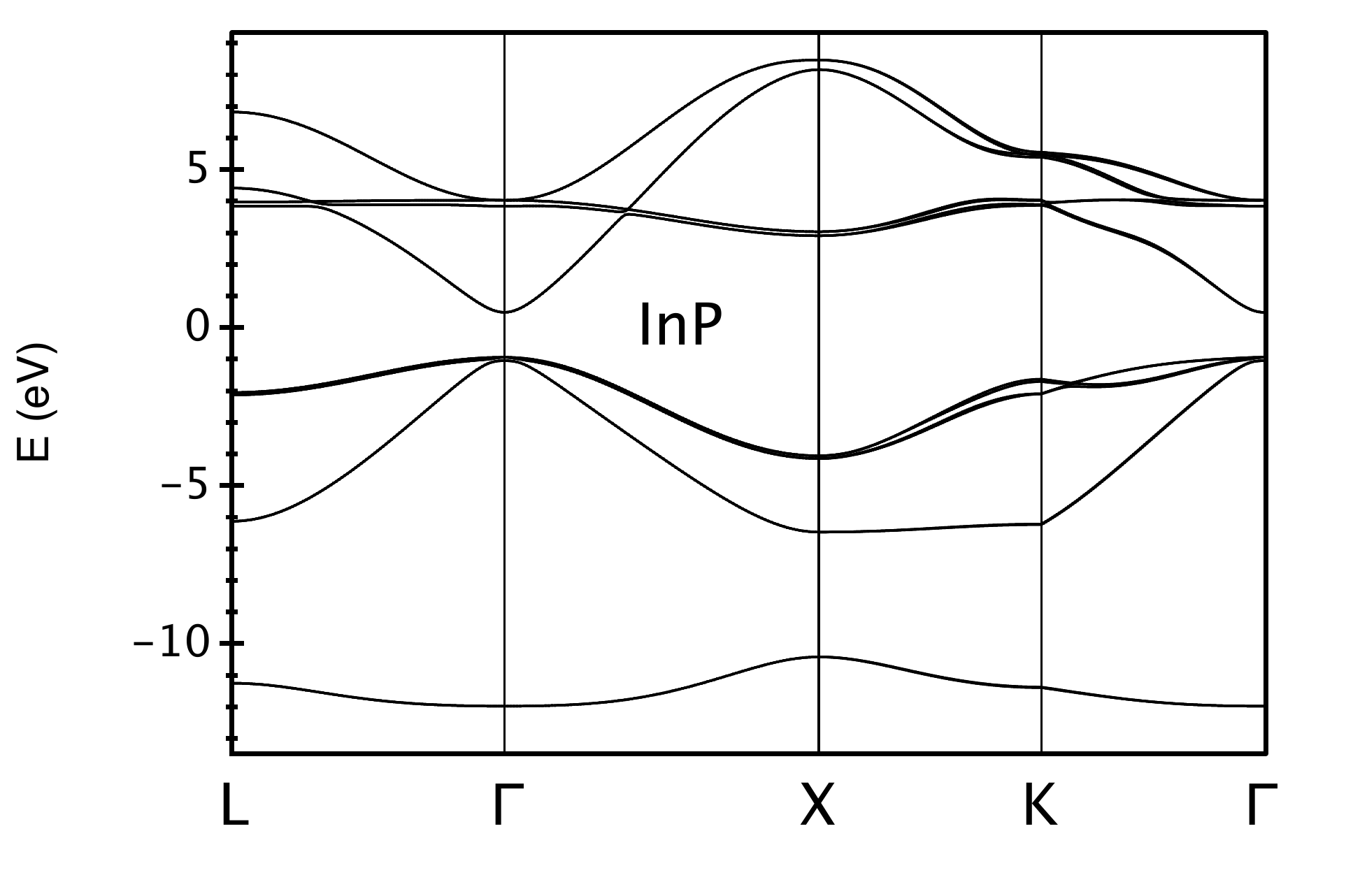}
\includegraphics[width=0.3\columnwidth]{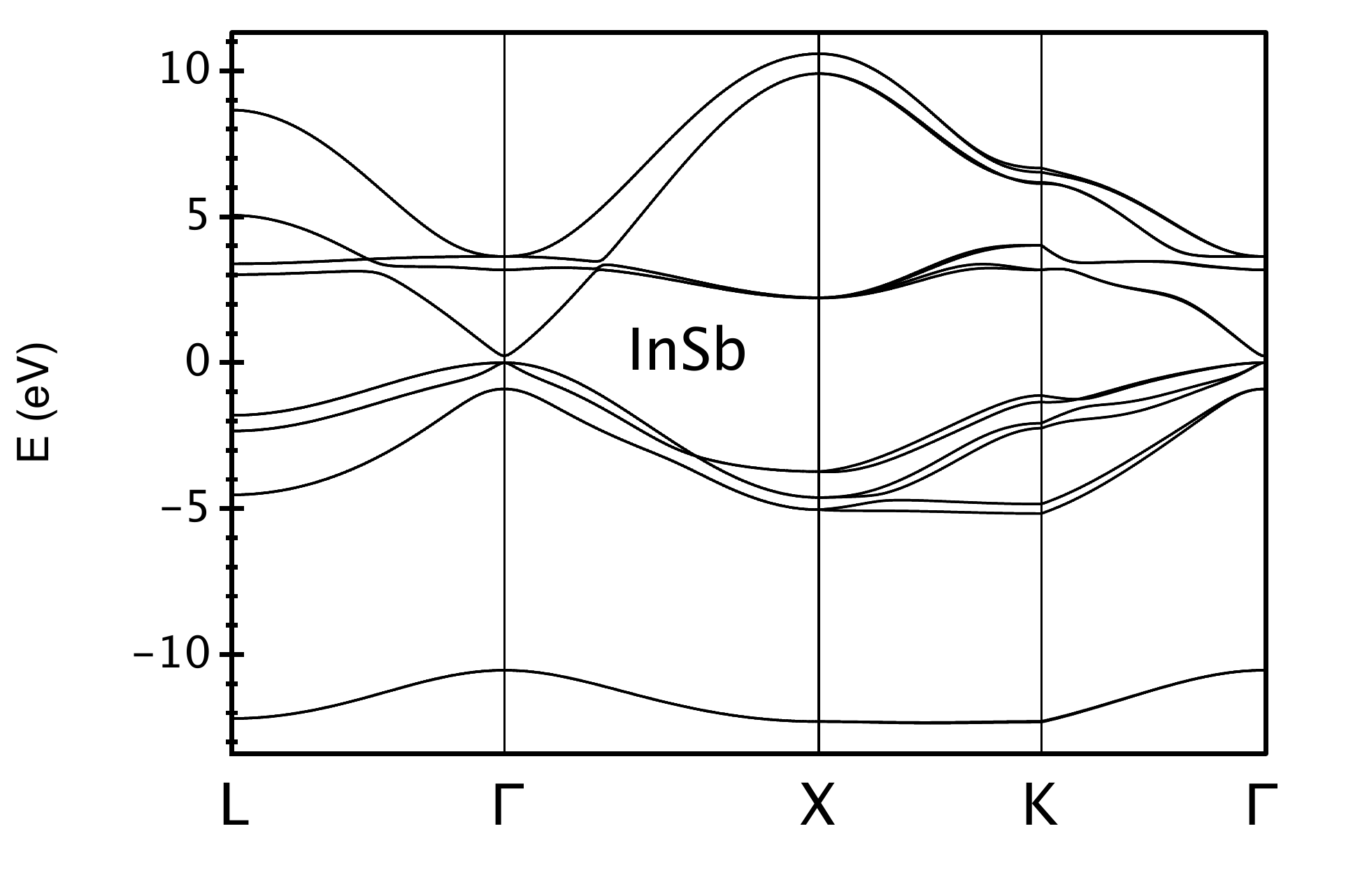}

\caption{Band diagrams using the parameters in Table \ref{tab:akpparams}.
Because the atomistic model is derived from a model that is perturbative in $\mathbf k$, it is inaccurate for large $\mathbf k$ and spurious solutions can cross the gap.
In a real-space formulation such spurious solutions will contribute to, and thus spoil, a numerical solution.
It is therefore important to examine the entire Brillouin zone associated with the computational grid (or the crystal lattice in the atomistic limit).
No gap-crossing states are seen for the parameters given in Table \ref{tab:akpparams}. InAs required an adjustment of the material parameters to avoid gap-crossing states (see Sect. \ref{sect:parameterFitting} ).
}
\label{fig:bands}
\end{figure}

The band structures resulting from our numerical fits are shown in Fig. \ref{fig:bands}. 
Since the atomistic model is derived from a continuum $k\cdot p$ model that is perturbative in $\mathbf k$, our results are accurate for small $\mathbf k$ and all materials appear to have a direct gap. 
Perturbative $k\cdot p$ models eventually break down at large $\mathbf k$, which can result in spurious solutions that cross the gap. 
When working in a plane wave basis these spurious solutions may be avoided by simply restricting the values of $\mathbf k$, however this cannot be done in a real-space formulation.
Spurious gap-crossing states can be eliminated by modifying the basis\cite{Foreman.prb.1997,Foreman.prb.2007}, choosing different material parameters\cite{Cartoixa.jap.2003}, or altering the differencing scheme to include higher powers of $\mathbf k$ that push the spurious solutions out of the gap\cite{Holm.jap.2002}.
The threat of gap-crossing bands is greater in the atomistic limit due to the larger 
({\it{computational}})
Brillouin zone associated with the smaller computational grid, providing more space for the bands to turn over and cross the gap.
We show energies throughout the entire Brillouin zone to demonstrate that our parameterization does not produce spurious gap-crossing states, and the model is suitable for use in a real-space formulation.
Only InAs required modifications to the parameters to suppress spurious solutions, as described above.

\begin{center}
\begin{table*}[ht]
\begin{tabular}{| l || c| c| c| c| c| c| c| c| c|}
 \hline
 \hline
{parameter} & AlP & GaP & InP & AlAs & GaAs &InAs & AlSb & GaSb & InSb \\
\hline
$a_{latt}~(\rm \AA)$ \footnotemark[1]& 5.4584 & 5.4417 & 5.8613 & 5.6524 & 5.6416 & 6.0501 & 6.1277 & 6.0817 & 6.4690 \\ 
$E_{6v}~(\rm eV)$ \footnotemark[2]& -11.21 &-12.14 &-11.04 &-11.73 &$-12.9 \footnotemark[3] $ &-11.53 &-10.62 &-11.47 & -10.54 \\
$E_{7v}~(\rm eV)$ \footnotemark[1] & -0.07 & -0.08 & -0.108 & -0.28 & -0.341 & -0.45\footnotemark[4] & -0.676 & -0.76 & -0.81 \\
$E_{8v}~(\rm eV)$ & $0~ $ &$0~ $ &$0~ $ &$0~ $ &$0~ $ & $0~ $ &$0~ $ &$0~ $ &$0~ $ \\
$E_{6c}~(\rm eV)$ \footnotemark[1] & 3.63 & 2.886 & 1.423 & 3.099 & 1.519 & 0.417 & 2.386 & 0.812 & 0.235 
 \\
$E_{7c}~(\rm eV)$ \footnotemark[5] & 4.78 & 4.38 & 4.78 & 4.55 & 4.488 \footnotemark[3] & 4.858\footnotemark[4] & 3.53 & 3.11 & 3.18 
 \\
$E_{8c}~(\rm eV)$ \footnotemark[5]&4.82 &4.47 &4.97 &4.70 & 4.659 \footnotemark[3]  &4.79 \footnotemark[4] &3.77 & 3.44 &3.64 \\
$m^*/m_0~ (1)$ \footnotemark[1]& 0.22 & 0.13 & 0.0795 & 0.15 & 0.067 & 0.026 & 0.14 & 0.039 & 0.0135 \\
$g^*/g_0~ (1)$ \footnotemark[5]& 1.92 & 1.9 & 1.26 & 1.52 & -0.44 & -14.9 & 0.84 & -9.2 & -51.6 \\ 
$\gamma_1^L~ (1)$ \footnotemark[1]& 3.35 & 4.05 & 5.08 & 3.76 & 6.98 & 20 & 5.18 & 13.4 & 34.8 \\
$\gamma_2^L~ (1)$ \footnotemark[1]& 0.714 & 0.49 & 1.6 & 0.82 & 2.06 & 7.5 \footnotemark[6] & 1.19 & 4.7 & 15.5 \\
$\gamma_3^L~ (1)$ \footnotemark[1]& 1.23 & 1.25 & 2.1 & 1.42 & 2.93 & 9.2 & 1.97 & 6.0 & 16.5 \\ 
$\gamma_c~(\rm eV \AA^3)$ \footnotemark[5]& 2.1 & -2.4 & -8.4 & 11.4 & 25.0 & 40.5 & 40.9 & 185.0 & 226.0 \\ 
\hline
\end{tabular}
\footnotetext[1]{ Ref. \onlinecite{Vurgaftman.jap.2001}, except where noted for InAs.}
\footnotetext[2]{ Ref. \onlinecite{Malone.jpcm.2013}, except for InAs.}
\footnotetext[3]{ Ref. \onlinecite{Aspnes.prb.1973}.}
\footnotetext[4]{Modified to fit $g^*$ and $\gamma_c$.}
\footnotetext[5]{ Ref. \onlinecite{Jancu.prb.2005} except where noted for GaAs and InAs.}
\footnotetext[6]{Modified to avoid gap-crossing bands at large $\mathbf k$.}
\caption {Target material parameters.
These are the physical material parameters which the atomistic model parameters were adjusted to fit. 
 The values of $g^*/g_0$ and $\gamma_c$ from Ref. \onlinecite{Jancu.prb.2005} are experimental values when available, and theoretical values when no measurements were available.}
\label{tab:target-params}
\end{table*}
\end{center}

\def\mymathhyphen{{\hbox{-}}}
\begin{center}
\begin{table*}[ht]
\begin{tabular}{| l || c| c| c| c| c| c| c| c| c|}
 \hline
 \hline
{parameter} & AlP & GaP & InP & AlAs & GaAs &InAs & AlSb & GaSb & InSb \\
\hline
 $E_{c1}$ (eV)& 1.560633 & 0.516670 & -1.501874 & 1.387058 & 0.675926 & -0.354316 & -1.094007 & -0.568950 & -11.520260 \\ 
 $E_{c2}$ (eV)& 5.236823 & 4.669935 & 3.360307 & 4.483957 & 2.273318 & 1.099526 & 4.568593 & 1.934999 & -0.968967 \\ 
 $E_{v1}$ (eV)& -0.505855 & -0.593772 & -0.604112 & -0.302607 & -0.751431 & -0.308368 & -0.328926 & -0.434964 & -0.310302 \\ 
 $E_{v2}$ (eV)& 0.645350 & 0.824859 & 0.809339 & 0.347832 & 1.162880 & 0.353026 & 0.400082 & 0.591758 & 3.264512 \\ 
 $E_{s1}$ (eV)& -0.620739 & -0.597985 & -0.591704 & -0.836975 & -1.198921 & -0.788711 & -1.423545 & -1.490988 & -1.213870 \\ 
 $E_{s2}$ (eV)& 0.650234 & 0.689072 & 0.498932 & 0.452200 & 1.098370 & -0.056631 & 0.578701 & 0.557782 & 2.808081 \\
$F_a$ (1)& -1.378388 & -1.376856 & -1.308404 & -1.450082 & -1.435248 & -1.390038 & -1.401502 & -1.411955 & -0.312572 \\ 
 $\gamma_{a1}$ (1)& -0.571909 & -0.514787 & -0.671323 & -0.609904 & -0.554424 & -0.731854 & -0.569585 & -0.498025 & -0.117695 \\ 
 $\gamma_{a2}$ (1)& 0.118013 & -0.733646 & -0.582392 & -0.292962 & -0.607743 & 0.572230 & 0.201367 & -0.087151 & -1.036618 \\ 
 $\gamma_{a3}$ (1)& 0.139473 & -0.334737 & -0.221699 & -0.010489 & -0.113528 & 0.633312 & 0.260787 & 0.246206 & -0.582445 \\ 
 $P_a$ ($ {\rm eV \,\AA}$)& 9.325446 & 10.188130 & 8.866846 & 10.192099 & 10.396101 & 8.791500 & 9.428844 & 10.123984 & 9.703173 \\ 
 $Q_a$ ($ {\rm eV\,\AA}$)& -7.062907 & -5.962138 & -5.085142 & -6.279963 & -6.139187 & -10.995692 & -7.559866 & -7.470334 & -4.704146 \\ 
\hline
\end{tabular}
\caption {Empirical atomistic $k.p$ theory parameters fit to the target parameters in Table I}
\label{tab:akpparams}
\end{table*}
\end{center}

\section{Conclusion}
\label{sect:conclusion}
We have demonstrated how to construct an atomistic $k\cdot p$ theory with finite differences on a grid matched to the crystal lattice.
Taking the atomistic limit of $k\cdot p$ theory in a straight-forward way results in a non-Hermitian Hamiltonian, which is seen to be related to the well known fact that multiplying a difference operator by a spatially varying coefficient leads to non-Hermiticity.
A more careful treatment shows the problem may be remedied by using the finite volume method, starting with an inversion symmetric Bloch basis, or by deforming the computational cells to generalized Voronoi cells.
The use of symmetric Bloch functions does not limit one to the symmetric approximation since the atomistic envelope functions themselves vary within the unit cell even at $\mathbf k= 0$.
The use of inversion symmetric Bloch functions and generalized Voronoi cells solve the Hermiticity problem, but are not applicable to heterojunctions.
As a result these approaches can be used on systems such as bulk materials, bulk materials with impurities or applied potentials, or nanocrystals with a vacuum barrier.
The finite volume method can be used in the presence of heterojunctions.

The atomistic limit of a simple four-band $k\cdot p$ model exactly reproduces the four-band tight-binding model, provided we include spherically symmetric remote band contributions for both the conduction and valence band, and the atomistic momentum matrix elements are different on different atoms (for zincblende).
In order to have different momentum matrix elements that make the model exactly match the tight-binding model requires the use of generalized Voronoi cells to to symmetrize the momentum matrix elements without making them all equal.
The atomistic limit of the widely used eight-band model results reproduces effective masses, g-factors, and Dresselhaus spin splittings of III-V materials.
The fits are exact for most materials, with the exception of InAs for which it was necessary to increase the spin-orbit coupling.
This may be due to an insufficient number of bands in the model, or due to uncertainties in the experimental values.

The particular implementation presented in Sect. \ref{sect:eight-bandModel} is by no means unique, and different atomistic models are possible depending on the choice of Bloch basis (inversion symmetric or not), the differencing scheme, and whether or not remote band contributions are included.
In addition, different fitting procedures may be used.
For example, the higher lying band energies could be left as free parameters adjusted to fit the band structure to other criteria such as charge asymmetry.
We have chosen to exactly fit all zone center energies, the zone center effective masses for the bottom of the conduction and top of the valence bands, as well as conduction g-factors and Dresselhaus spin splittings, since these quantities are the most important for electronic states of impurities and nanostructures.

An interesting property of these models is that the envelope functions have momenta outside the first Brillouin zone, a feature shared with the Burt-Foreman\cite{Burt.sst.1988a,Foreman.prb.1996} approach to dealing with heterojunctions.
Since our model is constructed in real space there is not a clearly defined separation between the wave function components that are associated with Bloch functions and those that are not, while the Burt-Foreman approach has a clear distinction between Bloch and envelope functions in $k$-space.

As seen from the fit to III-V materials, the atomistic envelope theory can reproduce the effective masses of the bands near the gap.
This is in contrast to tight-binding models which can give incorrect effective masses\cite{Boykin.prb.1998}.
The four-band model with only spherically symmetric remote band contributions illustrates how a nearest neighbor tight-binding model fails to reproduce the correct cubic band warping of the valence band.
In contrast, the atomistic Kane model gives the correct effective masses because it contains next nearest neighbor couplings via the Luttinger parameters.

There are many potential applications of this method to the electronic properties of impurity states, alloys, and polytypes\cite{De.prb.2010}. 
For sufficiently small nanoparticles we expect atomistic $k\cdot p$ theory to improve the description of the electronic structure, compared with continuum $k\cdot p$-theory. In particular nanoparticles with an irregular surface necessitate an atomistic description. It would be interesting to test how well our new method can describe structural defects such as dislocations, twin planes and stacking defects. Such systems cannot be easily treated in continuum k.p-theory. Atomistic $k\cdot p$ theory will also allow strain effects to be directly modelled in terms of atomic positions, a task which is difficult in both tight-binding and pseudopotential methods.
Finally, atomistic $k\cdot p$ theory has the unique feature that it allows the combination of atomistic and continuum models in the same system to facilitate multiscale modeling since the grid can be highly non-uniform.
One could use a rectilinear grid in "large" regions described by a continuum model and an atomistic grid in the regions requiring atomistic precision.
The differencing operators in the regions where the rectilinear and atomistic grids meet would be peculiar to the details of the grid used, but would be well defined. 
Multiscale modeling will dramatically reduce the computing time of atomistic k.p-theory compared with other atomistic models, while keeping atomistic accuracy where it is necessary.
\\
{\bf Acknowledgements}~
M.-E. P. acknowledges the support of the Swedish Research Council (VR).

 \section{Appendix: Bloch Basis States}
The Bloch state basis for the eight-band model is
\begin{align}
u_1=u^{\Gamma_6}_{-1/2} &= |S \downarrow \rangle \nonumber \\
u_2=u^{\Gamma_6}_{+1/2} &= |S \uparrow \rangle \nonumber \\
u_3=u^{\Gamma_8}_{+1/2} &= \frac{-i}{\sqrt 6} |(X+ i Y)\downarrow \rangle + i\sqrt{\frac{2}{3}}|Z\uparrow\rangle \nonumber \\
u_4=u^{\Gamma_8}_{+3/2} &= \frac{i}{\sqrt 2} |(X+ i Y)\uparrow \rangle \nonumber \\
u_5=u^{\Gamma_8}_{-3/2} &= \frac{-i}{\sqrt 2} |(X- i Y)\downarrow \rangle \nonumber \\
u_6=u^{\Gamma_8}_{-1/2} &= \frac{i}{\sqrt 6} |(X- i Y)\uparrow \rangle + i\sqrt{\frac{2}{3}}|Z\downarrow\rangle \nonumber \\
u_7=u^{\Gamma_7}_{-1/2} &= \frac{-i}{\sqrt 3} |(X- i Y)\uparrow \rangle + \frac{i}{\sqrt{3}}|Z\downarrow\rangle \nonumber \\
u_8=u^{\Gamma_7}_{+1/2} &= \frac{-i}{\sqrt 3} |(X+ i Y)\downarrow \rangle -\frac{i}{\sqrt{3}}|Z\uparrow\rangle \nonumber 
\end{align}
where the ordering of states is the same as for the Hamiltonian in Eq. \ref{eq:Bahder}.



\begin{thebibliography}{57}
\expandafter\ifx\csname natexlab\endcsname\relax\def\natexlab#1{#1}\fi
\expandafter\ifx\csname bibnamefont\endcsname\relax
  \def\bibnamefont#1{#1}\fi
\expandafter\ifx\csname bibfnamefont\endcsname\relax
  \def\bibfnamefont#1{#1}\fi
\expandafter\ifx\csname citenamefont\endcsname\relax
  \def\citenamefont#1{#1}\fi
\expandafter\ifx\csname url\endcsname\relax
  \def\url#1{\texttt{#1}}\fi
\expandafter\ifx\csname urlprefix\endcsname\relax\def\urlprefix{URL }\fi
\providecommand{\bibinfo}[2]{#2}
\providecommand{\eprint}[2][]{\url{#2}}

\bibitem[{\citenamefont{Cohen and Bergstresser}(1966)}]{Cohen.pr.1966}
\bibinfo{author}{\bibfnamefont{M.~L.} \bibnamefont{Cohen}} \bibnamefont{and}
  \bibinfo{author}{\bibfnamefont{T.~K.} \bibnamefont{Bergstresser}},
  \bibinfo{journal}{Phys. Rev.} \textbf{\bibinfo{volume}{141}},
  \bibinfo{pages}{789} (\bibinfo{year}{1966}),
  \urlprefix\url{http://dx.doi.org/10.1103/PhysRev.141.789}.

\bibitem[{\citenamefont{Chelikowsky and Cohen}(1976)}]{Chelikowsky.prb.1976}
\bibinfo{author}{\bibfnamefont{J.~R.} \bibnamefont{Chelikowsky}}
  \bibnamefont{and} \bibinfo{author}{\bibfnamefont{M.~L.} \bibnamefont{Cohen}},
  \bibinfo{journal}{Phys. Rev. B} \textbf{\bibinfo{volume}{14}},
  \bibinfo{pages}{556} (\bibinfo{year}{1976}),
  \urlprefix\url{http://dx.doi.org/10.1103/PhysRev.141.789}.

\bibitem[{\citenamefont{Vogl et~al.}(1983)\citenamefont{Vogl, Hjalmarson, and
  Dow}}]{Vogl.jpcs.1983}
\bibinfo{author}{\bibfnamefont{P.}~\bibnamefont{Vogl}},
  \bibinfo{author}{\bibfnamefont{H.~P.} \bibnamefont{Hjalmarson}},
  \bibnamefont{and} \bibinfo{author}{\bibfnamefont{J.~D.} \bibnamefont{Dow}},
  \bibinfo{journal}{J. Phys. Chem. Solids} \textbf{\bibinfo{volume}{44}},
  \bibinfo{pages}{365 } (\bibinfo{year}{1983}), ISSN \bibinfo{issn}{0022-3697},
  \urlprefix\url{http://dx.doi.org/10.1016/0022-3697(83)90064-1}.

\bibitem[{\citenamefont{Jancu et~al.}(1998)\citenamefont{Jancu, Scholz,
  Beltram, and Bassani}}]{Jancu.prb.1998}
\bibinfo{author}{\bibfnamefont{J.-M.} \bibnamefont{Jancu}},
  \bibinfo{author}{\bibfnamefont{R.}~\bibnamefont{Scholz}},
  \bibinfo{author}{\bibfnamefont{F.}~\bibnamefont{Beltram}}, \bibnamefont{and}
  \bibinfo{author}{\bibfnamefont{F.}~\bibnamefont{Bassani}},
  \bibinfo{journal}{Phys. Rev. B} \textbf{\bibinfo{volume}{57}},
  \bibinfo{pages}{6493} (\bibinfo{year}{1998}),
  \urlprefix\url{http://link.aps.org/doi/10.1103/PhysRevB.57.6493}.

\bibitem[{\citenamefont{Klimeck
  et~al.}(2000{\natexlab{a}})\citenamefont{Klimeck, Bowen, Boykin, and
  Cwik}}]{Klimeck.sm.2000b}
\bibinfo{author}{\bibfnamefont{G.}~\bibnamefont{Klimeck}},
  \bibinfo{author}{\bibfnamefont{R.~C.} \bibnamefont{Bowen}},
  \bibinfo{author}{\bibfnamefont{T.~B.} \bibnamefont{Boykin}},
  \bibnamefont{and} \bibinfo{author}{\bibfnamefont{T.~A.} \bibnamefont{Cwik}},
  \bibinfo{journal}{Superlattices Microstruct.} \textbf{\bibinfo{volume}{27}},
  \bibinfo{pages}{519 } (\bibinfo{year}{2000}{\natexlab{a}}), ISSN
  \bibinfo{issn}{0749-6036},
  \urlprefix\url{http://dx.doi.org/10.1006/spmi.2000.0862}.

\bibitem[{\citenamefont{Jancu et~al.}(2005)\citenamefont{Jancu, Scholz,
  de~Andrada~e Silva, and La~Rocca}}]{Jancu.prb.2005}
\bibinfo{author}{\bibfnamefont{J.-M.} \bibnamefont{Jancu}},
  \bibinfo{author}{\bibfnamefont{R.}~\bibnamefont{Scholz}},
  \bibinfo{author}{\bibfnamefont{E.~A.} \bibnamefont{de~Andrada~e Silva}},
  \bibnamefont{and} \bibinfo{author}{\bibfnamefont{G.~C.}
  \bibnamefont{La~Rocca}}, \bibinfo{journal}{Phys. Rev. B}
  \textbf{\bibinfo{volume}{72}}, \bibinfo{pages}{193201}
  (\bibinfo{year}{2005}),
  \urlprefix\url{http://dx.doi.org/10.1103/PhysRevB.72.193201}.

\bibitem[{\citenamefont{Voon and Willatzen}(2009)}]{Voon.book.2009}
\bibinfo{author}{\bibfnamefont{L.~L.~Y.} \bibnamefont{Voon}} \bibnamefont{and}
  \bibinfo{author}{\bibfnamefont{M.}~\bibnamefont{Willatzen}},
  \emph{\bibinfo{title}{The k.p Method: Electronic Properties of
  Semiconductors}} (\bibinfo{publisher}{Springer}, \bibinfo{year}{2009}).

\bibitem[{\citenamefont{Wang and Zunger}(1996)}]{Wang.prb.1996b}
\bibinfo{author}{\bibfnamefont{L.}~\bibnamefont{Wang}} \bibnamefont{and}
  \bibinfo{author}{\bibfnamefont{A.}~\bibnamefont{Zunger}},
  \bibinfo{journal}{Phys. Rev. B} \textbf{\bibinfo{volume}{54}},
  \bibinfo{pages}{11417} (\bibinfo{year}{1996}),
  \urlprefix\url{http://dx.doi.org/10.1103/PhysRevB.54.11417}.

\bibitem[{\citenamefont{Zunger}(2001)}]{Zunger.pss.2001}
\bibinfo{author}{\bibfnamefont{A.}~\bibnamefont{Zunger}},
  \bibinfo{journal}{physica status solidi (b)} \textbf{\bibinfo{volume}{224}},
  \bibinfo{pages}{727} (\bibinfo{year}{2001}), ISSN \bibinfo{issn}{1521-3951},
  \urlprefix\url{http://dx.doi.org/10.1002/(SICI)1521-3951(200104)224:3<727::AID-PSSB727>3.0.CO;2-9}.

\bibitem[{\citenamefont{Williamson and Zunger}(1999)}]{Williamson.prb.1999}
\bibinfo{author}{\bibfnamefont{A.~J.} \bibnamefont{Williamson}}
  \bibnamefont{and} \bibinfo{author}{\bibfnamefont{A.}~\bibnamefont{Zunger}},
  \bibinfo{journal}{Phys. Rev. B} \textbf{\bibinfo{volume}{59}},
  \bibinfo{pages}{15819} (\bibinfo{year}{1999}),
  \urlprefix\url{http://dx.doi.org/10.1103/PhysRevB.59.15819}.

\bibitem[{\citenamefont{Saito et~al.}(1998)\citenamefont{Saito, Schulman, and
  Arakawa}}]{Saito.prb.1998}
\bibinfo{author}{\bibfnamefont{T.}~\bibnamefont{Saito}},
  \bibinfo{author}{\bibfnamefont{J.~N.} \bibnamefont{Schulman}},
  \bibnamefont{and} \bibinfo{author}{\bibfnamefont{Y.}~\bibnamefont{Arakawa}},
  \bibinfo{journal}{Phys. Rev. B} \textbf{\bibinfo{volume}{57}},
  \bibinfo{pages}{13016} (\bibinfo{year}{1998}),
  \urlprefix\url{http://link.aps.org/doi/10.1103/PhysRevB.57.13016}.

\bibitem[{\citenamefont{Jask\'olski et~al.}(2006)\citenamefont{Jask\'olski,
  Zieli\ifmmode~\acute{n}\else \'{n}\fi{}ski, Bryant, and
  Aizpurua}}]{Jaskolski.prb.2006}
\bibinfo{author}{\bibfnamefont{W.}~\bibnamefont{Jask\'olski}},
  \bibinfo{author}{\bibfnamefont{M.}~\bibnamefont{Zieli\ifmmode~\acute{n}\else
  \'{n}\fi{}ski}}, \bibinfo{author}{\bibfnamefont{G.~W.} \bibnamefont{Bryant}},
  \bibnamefont{and} \bibinfo{author}{\bibfnamefont{J.}~\bibnamefont{Aizpurua}},
  \bibinfo{journal}{Phys. Rev. B} \textbf{\bibinfo{volume}{74}},
  \bibinfo{pages}{195339} (\bibinfo{year}{2006}),
  \urlprefix\url{http://dx.doi.org/10.1103/PhysRevB.74.195339}.

\bibitem[{\citenamefont{Grundmann et~al.}(1995)\citenamefont{Grundmann, Stier,
  and Bimberg}}]{Grundmann.prb.1995}
\bibinfo{author}{\bibfnamefont{M.}~\bibnamefont{Grundmann}},
  \bibinfo{author}{\bibfnamefont{O.}~\bibnamefont{Stier}}, \bibnamefont{and}
  \bibinfo{author}{\bibfnamefont{D.}~\bibnamefont{Bimberg}},
  \bibinfo{journal}{Phys. Rev. B} \textbf{\bibinfo{volume}{52}},
  \bibinfo{pages}{11969} (\bibinfo{year}{1995}),
  \urlprefix\url{http://dx.doi.org/10.1103/PhysRevB.52.11969}.

\bibitem[{\citenamefont{Cusack et~al.}(1996)\citenamefont{Cusack, Briddon, and
  Jaros}}]{Cusack.prb.1996}
\bibinfo{author}{\bibfnamefont{M.~A.} \bibnamefont{Cusack}},
  \bibinfo{author}{\bibfnamefont{P.~R.} \bibnamefont{Briddon}},
  \bibnamefont{and} \bibinfo{author}{\bibfnamefont{M.}~\bibnamefont{Jaros}},
  \bibinfo{journal}{Phys. Rev. B} \textbf{\bibinfo{volume}{54}},
  \bibinfo{pages}{R2300} (\bibinfo{year}{1996}),
  \urlprefix\url{http://dx.doi.org/10.1103/PhysRevB.54.R2300}.

\bibitem[{\citenamefont{Jiang and Singh}(1997)}]{Jiang.prb.1997}
\bibinfo{author}{\bibfnamefont{H.}~\bibnamefont{Jiang}} \bibnamefont{and}
  \bibinfo{author}{\bibfnamefont{J.}~\bibnamefont{Singh}},
  \bibinfo{journal}{Phys. Rev. B} \textbf{\bibinfo{volume}{56}},
  \bibinfo{pages}{4696} (\bibinfo{year}{1997}),
  \urlprefix\url{http://dx.doi.org/10.1103/PhysRevB.56.4696}.

\bibitem[{\citenamefont{Pryor}(1998)}]{Pryor.prb.1998}
\bibinfo{author}{\bibfnamefont{C.}~\bibnamefont{Pryor}},
  \bibinfo{journal}{Phys. Rev. B} \textbf{\bibinfo{volume}{57}},
  \bibinfo{pages}{7190} (\bibinfo{year}{1998}),
  \urlprefix\url{http://dx.doi.org/10.1103/PhysRevB.57.7190}.

\bibitem[{\citenamefont{Goodwin et~al.}(1989)\citenamefont{Goodwin, Skinner,
  and Pettifor}}]{Goodwin.epl.1989}
\bibinfo{author}{\bibfnamefont{L.}~\bibnamefont{Goodwin}},
  \bibinfo{author}{\bibfnamefont{A.~J.} \bibnamefont{Skinner}},
  \bibnamefont{and} \bibinfo{author}{\bibfnamefont{D.~G.}
  \bibnamefont{Pettifor}}, \bibinfo{journal}{epl} \textbf{\bibinfo{volume}{9}},
  \bibinfo{pages}{701} (\bibinfo{year}{1989}),
  \urlprefix\url{http://stacks.iop.org/0295-5075/9/i=7/a=015}.

\bibitem[{\citenamefont{Klimeck
  et~al.}(2000{\natexlab{b}})\citenamefont{Klimeck, Bowen, Boykin,
  Salazar-Lazaro, Cwik, and Stoica}}]{Klimeck.sm.2000a}
\bibinfo{author}{\bibfnamefont{G.}~\bibnamefont{Klimeck}},
  \bibinfo{author}{\bibfnamefont{R.~C.} \bibnamefont{Bowen}},
  \bibinfo{author}{\bibfnamefont{T.~B.} \bibnamefont{Boykin}},
  \bibinfo{author}{\bibfnamefont{C.}~\bibnamefont{Salazar-Lazaro}},
  \bibinfo{author}{\bibfnamefont{T.~A.} \bibnamefont{Cwik}}, \bibnamefont{and}
  \bibinfo{author}{\bibfnamefont{A.}~\bibnamefont{Stoica}},
  \bibinfo{journal}{Superlattices Microstruct.} \textbf{\bibinfo{volume}{27}},
  \bibinfo{pages}{77} (\bibinfo{year}{2000}{\natexlab{b}}),
  \urlprefix\url{http://dx.doi.org/10.1006/spmi.1999.0797}.

\bibitem[{\citenamefont{Kim et~al.}(1998)\citenamefont{Kim, Wang, and
  Zunger}}]{Kim.prb.1998}
\bibinfo{author}{\bibfnamefont{J.}~\bibnamefont{Kim}},
  \bibinfo{author}{\bibfnamefont{L.-W.} \bibnamefont{Wang}}, \bibnamefont{and}
  \bibinfo{author}{\bibfnamefont{A.}~\bibnamefont{Zunger}},
  \bibinfo{journal}{Phys. Rev. B} \textbf{\bibinfo{volume}{57}},
  \bibinfo{pages}{R9408} (\bibinfo{year}{1998}),
  \urlprefix\url{http://dx.doi.org/10.1103/PhysRevB.57.R9408}.

\bibitem[{\citenamefont{Cardona and Pollak}(1966)}]{Cardona.pr.1966}
\bibinfo{author}{\bibfnamefont{M.}~\bibnamefont{Cardona}} \bibnamefont{and}
  \bibinfo{author}{\bibfnamefont{F.~H.} \bibnamefont{Pollak}},
  \bibinfo{journal}{Phys. Rev.} \textbf{\bibinfo{volume}{142}},
  \bibinfo{pages}{530} (\bibinfo{year}{1966}),
  \urlprefix\url{http://dx.doi.org/10.1103/PhysRev.142.530}.

\bibitem[{\citenamefont{Fraj et~al.}(2008)\citenamefont{Fraj, Saidi, Radhia,
  and Boujdaria}}]{Fraj.sst.2008}
\bibinfo{author}{\bibfnamefont{N.}~\bibnamefont{Fraj}},
  \bibinfo{author}{\bibfnamefont{I.}~\bibnamefont{Saidi}},
  \bibinfo{author}{\bibfnamefont{S.~B.} \bibnamefont{Radhia}},
  \bibnamefont{and}
  \bibinfo{author}{\bibfnamefont{K.}~\bibnamefont{Boujdaria}},
  \bibinfo{journal}{Semiconductor Science and Technology}
  \textbf{\bibinfo{volume}{23}}, \bibinfo{pages}{085006}
  (\bibinfo{year}{2008}),
  \urlprefix\url{http://dx.doi.org/10.1088/0268-1242/23/8/085006}.

\bibitem[{\citenamefont{Fraj et~al.}(2007)\citenamefont{Fraj, Sa\"idi, Radhia,
  and Boujdaria}}]{Fraj.jap.2007}
\bibinfo{author}{\bibfnamefont{N.}~\bibnamefont{Fraj}},
  \bibinfo{author}{\bibfnamefont{I.}~\bibnamefont{Sa\"idi}},
  \bibinfo{author}{\bibfnamefont{S.~B.} \bibnamefont{Radhia}},
  \bibnamefont{and}
  \bibinfo{author}{\bibfnamefont{K.}~\bibnamefont{Boujdaria}},
  \bibinfo{journal}{J. Appl. Phys.} \textbf{\bibinfo{volume}{102}},
  \bibinfo{pages}{053703} (\bibinfo{year}{2007}),
  \urlprefix\url{http://dx.doi.org/10.1063/1.2773532}.

\bibitem[{\citenamefont{Richard et~al.}(2005)\citenamefont{Richard, Aniel, and
  Fishman}}]{Richard.prb.2005}
\bibinfo{author}{\bibfnamefont{S.}~\bibnamefont{Richard}},
  \bibinfo{author}{\bibfnamefont{F.}~\bibnamefont{Aniel}}, \bibnamefont{and}
  \bibinfo{author}{\bibfnamefont{G.}~\bibnamefont{Fishman}},
  \bibinfo{journal}{Phys. Rev. B} \textbf{\bibinfo{volume}{72}},
  \bibinfo{pages}{245316} (\bibinfo{year}{2005}),
  \urlprefix\url{http://dx.doi.org/10.1103/PhysRevB.72.245316}.

\bibitem[{\citenamefont{Saidi et~al.}(2008)\citenamefont{Saidi, Ben~Radhia, and
  Boujdaria}}]{Saidi.jap.2008}
\bibinfo{author}{\bibfnamefont{I.}~\bibnamefont{Saidi}},
  \bibinfo{author}{\bibfnamefont{S.}~\bibnamefont{Ben~Radhia}},
  \bibnamefont{and}
  \bibinfo{author}{\bibfnamefont{K.}~\bibnamefont{Boujdaria}},
  \bibinfo{journal}{Journal of Applied Physics} \textbf{\bibinfo{volume}{104}},
  \bibinfo{eid}{023706} (\bibinfo{year}{2008}),
  \urlprefix\url{http://dx.doi.org/10.1063/1.2957068}.

\bibitem[{\citenamefont{Pryor}(1991)}]{Pryor.prb.1991}
\bibinfo{author}{\bibfnamefont{C.}~\bibnamefont{Pryor}},
  \bibinfo{journal}{Phys. Rev. B} \textbf{\bibinfo{volume}{44}},
  \bibinfo{pages}{12912} (\bibinfo{year}{1991}),
  \urlprefix\url{http://dx.doi.org/10.1103/PhysRevB.44.12912}.

\bibitem[{\citenamefont{Pryor et~al.}(1997)\citenamefont{Pryor, Pistol, and
  Samuelson}}]{Pryor.prb.1997}
\bibinfo{author}{\bibfnamefont{C.}~\bibnamefont{Pryor}},
  \bibinfo{author}{\bibfnamefont{M.-E.} \bibnamefont{Pistol}},
  \bibnamefont{and}
  \bibinfo{author}{\bibfnamefont{L.}~\bibnamefont{Samuelson}},
  \bibinfo{journal}{Phys. Rev. B} \textbf{\bibinfo{volume}{56}},
  \bibinfo{pages}{10404} (\bibinfo{year}{1997}),
  \urlprefix\url{http://dx.doi.org/10.1103/PhysRevB.56.10404}.

\bibitem[{\citenamefont{Cullum and Willoughby}(1985)}]{Cullum.book.1985}
\bibinfo{author}{\bibfnamefont{J.~K.} \bibnamefont{Cullum}} \bibnamefont{and}
  \bibinfo{author}{\bibfnamefont{R.~A.} \bibnamefont{Willoughby}},
  \emph{\bibinfo{title}{Lanczos algorithms for large symmetric eigenvalue
  computations}}, vol. \bibinfo{volume}{1-2}
  (\bibinfo{publisher}{(Birkh\"{a}user, Boston)}, \bibinfo{year}{1985}).

\bibitem[{\citenamefont{Wang and Zunger}(1994)}]{Wang.jcp.1994}
\bibinfo{author}{\bibfnamefont{L.-W.} \bibnamefont{Wang}} \bibnamefont{and}
  \bibinfo{author}{\bibfnamefont{A.}~\bibnamefont{Zunger}},
  \bibinfo{journal}{J. Chem. Phys.} \textbf{\bibinfo{volume}{100}},
  \bibinfo{pages}{2394} (\bibinfo{year}{1994}),
  \urlprefix\url{http://dx.doi.org/10.1063/1.466486}.

\bibitem[{\citenamefont{Abramowitz and Stegun}(1964)}]{Abramowitz.book.1964}
\bibinfo{author}{\bibfnamefont{M.}~\bibnamefont{Abramowitz}} \bibnamefont{and}
  \bibinfo{author}{\bibfnamefont{I.~A.} \bibnamefont{Stegun}},
  \emph{\bibinfo{title}{Handbook of mathematical functions with formulas,
  graphs, and mathematical tables}}, vol.~\bibinfo{volume}{55} of
  \emph{\bibinfo{series}{National Bureau of Standards Applied Mathematics
  Series}} (\bibinfo{publisher}{For sale by the Superintendent of Documents,
  U.S. Government Printing Office, Washington, D.C.}, \bibinfo{year}{1964}).

\bibitem[{\citenamefont{Fornberg}(1988)}]{Fornberg.mc.1988}
\bibinfo{author}{\bibfnamefont{B.}~\bibnamefont{Fornberg}},
  \bibinfo{journal}{Math. of Comp.} \textbf{\bibinfo{volume}{51}},
  \bibinfo{pages}{699} (\bibinfo{year}{1988}),
  \urlprefix\url{http://dx.doi.org/10.1090/S0025-5718-1988-0935077-0}.

\bibitem[{\citenamefont{Lakin}(1986)}]{Lakin.ijnme.1986}
\bibinfo{author}{\bibfnamefont{W.~D.} \bibnamefont{Lakin}},
  \bibinfo{journal}{Internat. J. Numer.Methods Engrg.}
  \textbf{\bibinfo{volume}{23}}, \bibinfo{pages}{209} (\bibinfo{year}{1986}),
  \urlprefix\url{http://dx.doi.org/10.1002/nme.1620230205}.

\bibitem[{\citenamefont{Beck}(2000)}]{Beck.rmp.2000}
\bibinfo{author}{\bibfnamefont{T.}~\bibnamefont{Beck}}, \bibinfo{journal}{Rev.
  Mod. Phys.} \textbf{\bibinfo{volume}{72}}, \bibinfo{pages}{1041}
  (\bibinfo{year}{2000}),
  \urlprefix\url{http://dx.doi.org/10.1103/RevModPhys.72.1041}.

\bibitem[{\citenamefont{Cardona et~al.}(1988)\citenamefont{Cardona,
  Christensen, and Fasol}}]{Cardona.prb.1988}
\bibinfo{author}{\bibfnamefont{M.}~\bibnamefont{Cardona}},
  \bibinfo{author}{\bibfnamefont{N.~E.} \bibnamefont{Christensen}},
  \bibnamefont{and} \bibinfo{author}{\bibfnamefont{G.}~\bibnamefont{Fasol}},
  \bibinfo{journal}{Phys. Rev. B} \textbf{\bibinfo{volume}{38}},
  \bibinfo{pages}{1806} (\bibinfo{year}{1988}),
  \urlprefix\url{http://dx.doi.org/10.1103/PhysRevB.38.1806}.

\bibitem[{\citenamefont{Pfeffer and Zawadzki}(1996)}]{Pfeffer.prb.1996}
\bibinfo{author}{\bibfnamefont{P.}~\bibnamefont{Pfeffer}} \bibnamefont{and}
  \bibinfo{author}{\bibfnamefont{W.}~\bibnamefont{Zawadzki}},
  \bibinfo{journal}{Phys. Rev. B} \textbf{\bibinfo{volume}{53}},
  \bibinfo{pages}{12813} (\bibinfo{year}{1996}),
  \urlprefix\url{http://dx.doi.org/10.1103/PhysRevB.53.12813}.

\bibitem[{\citenamefont{Mishev}(1998)}]{Mishev.nmpde.1998}
\bibinfo{author}{\bibfnamefont{I.~D.} \bibnamefont{Mishev}},
  \bibinfo{journal}{Numerical Methods for Partial Differential Equations}
  \textbf{\bibinfo{volume}{14}}, \bibinfo{pages}{193} (\bibinfo{year}{1998}),
  ISSN \bibinfo{issn}{1098-2426},
  \urlprefix\url{http://dx.doi.org/10.1002/(SICI)1098-2426(199803)14:2<193::AID-NUM4>3.0.CO;2-J}.

\bibitem[{\citenamefont{Telea and van Wijk}(2001)}]{Telea.book.2011}
\bibinfo{author}{\bibfnamefont{A.}~\bibnamefont{Telea}} \bibnamefont{and}
  \bibinfo{author}{\bibfnamefont{J.}~\bibnamefont{van Wijk}}, in
  \emph{\bibinfo{booktitle}{Data Visualization 2001}}, edited by
  \bibinfo{editor}{\bibfnamefont{D.}~\bibnamefont{Ebert}},
  \bibinfo{editor}{\bibfnamefont{J.}~\bibnamefont{Favre}}, \bibnamefont{and}
  \bibinfo{editor}{\bibfnamefont{R.}~\bibnamefont{Peikert}}
  (\bibinfo{publisher}{Springer Vienna}, \bibinfo{year}{2001}), Eurographics,
  pp. \bibinfo{pages}{165--174}, ISBN \bibinfo{isbn}{978-3-211-83674-3},
  \urlprefix\url{http://dx.doi.org/10.1007/978-3-7091-6215-6_18}.

\bibitem[{\citenamefont{Chadi and Cohen}(1975)}]{Chadi.pss.1975}
\bibinfo{author}{\bibfnamefont{D.~J.} \bibnamefont{Chadi}} \bibnamefont{and}
  \bibinfo{author}{\bibfnamefont{M.~L.} \bibnamefont{Cohen}},
  \bibinfo{journal}{physica status solidi (b)} \textbf{\bibinfo{volume}{68}},
  \bibinfo{pages}{405} (\bibinfo{year}{1975}), ISSN \bibinfo{issn}{1521-3951},
  \urlprefix\url{http://dx.doi.org/10.1002/pssb.2220680140}.

\bibitem[{\citenamefont{Zhu and Kroemer}(1983)}]{Zhu.prb.1983}
\bibinfo{author}{\bibfnamefont{Q.}~\bibnamefont{Zhu}} \bibnamefont{and}
  \bibinfo{author}{\bibfnamefont{H.}~\bibnamefont{Kroemer}},
  \bibinfo{journal}{Phys. Rev. B} \textbf{\bibinfo{volume}{27}},
  \bibinfo{pages}{3519} (\bibinfo{year}{1983}),
  \urlprefix\url{http://dx.doi.org/10.1103/PhysRevB.27.3519}.

\bibitem[{\citenamefont{Birman and Solomjak}(1986)}]{Birman.book.1986}
\bibinfo{author}{\bibfnamefont{M.~S.} \bibnamefont{Birman}} \bibnamefont{and}
  \bibinfo{author}{\bibfnamefont{M.~Z.} \bibnamefont{Solomjak}},
  \emph{\bibinfo{title}{Spectral Theory of Self-Adjoint Operators in Hilbert
  Space}} (\bibinfo{publisher}{D. Reidel Publishing Company},
  \bibinfo{address}{Dordrecht}, \bibinfo{year}{1986}).

\bibitem[{\citenamefont{Einevoll et~al.}(1990)\citenamefont{Einevoll, Hemmer,
  and Thomsen}}]{Einevoll.prb.1990}
\bibinfo{author}{\bibfnamefont{G.~T.} \bibnamefont{Einevoll}},
  \bibinfo{author}{\bibfnamefont{P.~C.} \bibnamefont{Hemmer}},
  \bibnamefont{and} \bibinfo{author}{\bibfnamefont{J.}~\bibnamefont{Thomsen}},
  \bibinfo{journal}{Phys. Rev. B} \textbf{\bibinfo{volume}{42}},
  \bibinfo{pages}{3485} (\bibinfo{year}{1990}),
  \urlprefix\url{http://dx.doi.org/10.1103/PhysRevB.42.3485}.

\bibitem[{\citenamefont{Knabner and Angerman}(2003)}]{Knabner.book.2003}
\bibinfo{author}{\bibfnamefont{P.}~\bibnamefont{Knabner}} \bibnamefont{and}
  \bibinfo{author}{\bibfnamefont{L.}~\bibnamefont{Angerman}},
  \emph{\bibinfo{title}{Numerical Methods for Elliptic and Parabolic Partial
  Differential Equations}} (\bibinfo{publisher}{Springer},
  \bibinfo{year}{2003}).

\bibitem[{\citenamefont{Burt}(1988)}]{Burt.sst.1988a}
\bibinfo{author}{\bibfnamefont{M.~G.} \bibnamefont{Burt}},
  \bibinfo{journal}{Semicond. Sci. Tech.} \textbf{\bibinfo{volume}{3}},
  \bibinfo{pages}{739} (\bibinfo{year}{1988}),
  \urlprefix\url{http://dx.doi.org/10.1088/0268-1242/3/8/003}.

\bibitem[{\citenamefont{Foreman}(1996)}]{Foreman.prb.1996}
\bibinfo{author}{\bibfnamefont{B.~A.} \bibnamefont{Foreman}},
  \bibinfo{journal}{Phys. Rev. B} \textbf{\bibinfo{volume}{54}},
  \bibinfo{pages}{1909} (\bibinfo{year}{1996}),
  \urlprefix\url{http://dx.doi.org/10.1103/PhysRevB.54.1909}.

\bibitem[{\citenamefont{Pidgeon and Brown}(1966)}]{Pidgeon.pr.1966}
\bibinfo{author}{\bibfnamefont{C.}~\bibnamefont{Pidgeon}} \bibnamefont{and}
  \bibinfo{author}{\bibfnamefont{R.}~\bibnamefont{Brown}},
  \bibinfo{journal}{Phys. Rev.} \textbf{\bibinfo{volume}{146}},
  \bibinfo{pages}{575} (\bibinfo{year}{1966}),
  \urlprefix\url{http://dx.doi.org/10.1103/PhysRev.146.575}.

\bibitem[{\citenamefont{Kane}(1957)}]{Kane.jpcs.1957}
\bibinfo{author}{\bibfnamefont{E.~O.} \bibnamefont{Kane}}, \bibinfo{journal}{J.
  Phys. Chem. Solids} \textbf{\bibinfo{volume}{1}}, \bibinfo{pages}{249}
  (\bibinfo{year}{1957}),
  \urlprefix\url{http://dx.doi.org/10.1016/0022-3697(57)90013-6}.

\bibitem[{\citenamefont{Bahder}(1990)}]{Bahder.prb.1990}
\bibinfo{author}{\bibfnamefont{T.~B.} \bibnamefont{Bahder}},
  \bibinfo{journal}{Phys. Rev. B} \textbf{\bibinfo{volume}{41}},
  \bibinfo{pages}{11992} (\bibinfo{year}{1990}),
  \urlprefix\url{http://dx.doi.org/10.1103/PhysRevB.41.11992}.

\bibitem[{\citenamefont{Aspnes and Studna}(1973)}]{Aspnes.prb.1973}
\bibinfo{author}{\bibfnamefont{D.~E.} \bibnamefont{Aspnes}} \bibnamefont{and}
  \bibinfo{author}{\bibfnamefont{A.~A.} \bibnamefont{Studna}},
  \bibinfo{journal}{Phys. Rev. B} \textbf{\bibinfo{volume}{7}},
  \bibinfo{pages}{4605} (\bibinfo{year}{1973}),
  \urlprefix\url{http://dx.doi.org/10.1103/PhysRevB.7.4605}.

\bibitem[{\citenamefont{Hermann and Weisbuch}(1977)}]{Hermann.prb.1977}
\bibinfo{author}{\bibfnamefont{C.}~\bibnamefont{Hermann}} \bibnamefont{and}
  \bibinfo{author}{\bibfnamefont{C.}~\bibnamefont{Weisbuch}},
  \bibinfo{journal}{Phys. Rev. B} \textbf{\bibinfo{volume}{15}},
  \bibinfo{pages}{823} (\bibinfo{year}{1977}),
  \urlprefix\url{http://dx.doi.org/10.1103/PhysRevB.15.823}.

\bibitem[{\citenamefont{Roth et~al.}(1959)\citenamefont{Roth, Lax, and
  Zwerdling}}]{Roth.pr.1959}
\bibinfo{author}{\bibfnamefont{L.~M.} \bibnamefont{Roth}},
  \bibinfo{author}{\bibfnamefont{B.}~\bibnamefont{Lax}}, \bibnamefont{and}
  \bibinfo{author}{\bibfnamefont{S.}~\bibnamefont{Zwerdling}},
  \bibinfo{journal}{Phys. Rev.} \textbf{\bibinfo{volume}{114}},
  \bibinfo{pages}{90} (\bibinfo{year}{1959}),
  \urlprefix\url{http://dx.doi.org/10.1103/PhysRev.114.90}.

\bibitem[{\citenamefont{Vurgaftman et~al.}(2001)\citenamefont{Vurgaftman,
  Meyer, and Ram-Mohan}}]{Vurgaftman.jap.2001}
\bibinfo{author}{\bibfnamefont{I.}~\bibnamefont{Vurgaftman}},
  \bibinfo{author}{\bibfnamefont{J.}~\bibnamefont{Meyer}}, \bibnamefont{and}
  \bibinfo{author}{\bibfnamefont{L.}~\bibnamefont{Ram-Mohan}},
  \bibinfo{journal}{Journ. Appl. Phys.} \textbf{\bibinfo{volume}{89}},
  \bibinfo{pages}{5815} (\bibinfo{year}{2001}),
  \urlprefix\url{http://dx.doi.org/10.1063/1.1368156}.

\bibitem[{\citenamefont{Foreman}(1997)}]{Foreman.prb.1997}
\bibinfo{author}{\bibfnamefont{B.~A.} \bibnamefont{Foreman}},
  \bibinfo{journal}{Phys. Rev. B} \textbf{\bibinfo{volume}{56}},
  \bibinfo{pages}{R12748} (\bibinfo{year}{1997}),
  \urlprefix\url{http://dx.doi.org/10.1103/PhysRevB.56.R12748}.

\bibitem[{\citenamefont{Foreman}(2007)}]{Foreman.prb.2007}
\bibinfo{author}{\bibfnamefont{B.~A.} \bibnamefont{Foreman}},
  \bibinfo{journal}{Phys. Rev. B} \textbf{\bibinfo{volume}{75}},
  \bibinfo{pages}{235331} (\bibinfo{year}{2007}),
  \urlprefix\url{http://dx.doi.org/10.1103/PhysRevB.75.235331}.

\bibitem[{\citenamefont{Cartoixa}(2003)}]{Cartoixa.jap.2003}
\bibinfo{author}{\bibfnamefont{X.}~\bibnamefont{Cartoixa}},
  \bibinfo{journal}{J. Appl. Phys.} \textbf{\bibinfo{volume}{68}},
  \bibinfo{pages}{235319} (\bibinfo{year}{2003}),
  \urlprefix\url{http://dx.doi.org/10.1063/1.1555833}.

\bibitem[{\citenamefont{Holm et~al.}(2002)\citenamefont{Holm, Pistol, and
  Pryor}}]{Holm.jap.2002}
\bibinfo{author}{\bibfnamefont{M.}~\bibnamefont{Holm}},
  \bibinfo{author}{\bibfnamefont{M.-E.} \bibnamefont{Pistol}},
  \bibnamefont{and} \bibinfo{author}{\bibfnamefont{C.}~\bibnamefont{Pryor}},
  \bibinfo{journal}{J. Appl. Phys.} \textbf{\bibinfo{volume}{92}},
  \bibinfo{pages}{932} (\bibinfo{year}{2002}),
  \urlprefix\url{http://dx.doi.org/10.1063/1.1486021}.

\bibitem[{\citenamefont{Malone and Cohen}(2013)}]{Malone.jpcm.2013}
\bibinfo{author}{\bibfnamefont{B.~D.} \bibnamefont{Malone}} \bibnamefont{and}
  \bibinfo{author}{\bibfnamefont{M.~L.} \bibnamefont{Cohen}},
  \bibinfo{journal}{Journal of Physics: Condensed Matter}
  \textbf{\bibinfo{volume}{25}}, \bibinfo{pages}{105503}
  (\bibinfo{year}{2013}),
  \urlprefix\url{http://dx.doi.org/10.1088/0953-8984/25/10/105503}.

\bibitem[{\citenamefont{Boykin}(1998)}]{Boykin.prb.1998}
\bibinfo{author}{\bibfnamefont{T.~B.} \bibnamefont{Boykin}},
  \bibinfo{journal}{Phys. Rev. B} \textbf{\bibinfo{volume}{57}},
  \bibinfo{pages}{1620} (\bibinfo{year}{1998}),
  \urlprefix\url{http://dx.doi.org/10.1103/PhysRevB.57.1620}.

\bibitem[{\citenamefont{De and Pryor}(2010)}]{De.prb.2010}
\bibinfo{author}{\bibfnamefont{A.}~\bibnamefont{De}} \bibnamefont{and}
  \bibinfo{author}{\bibfnamefont{C.~E.} \bibnamefont{Pryor}},
  \bibinfo{journal}{Phys. Rev. B} \textbf{\bibinfo{volume}{81}},
  \bibinfo{pages}{155210} (\bibinfo{year}{2010}),
  \urlprefix\url{http://dx.doi.org/10.1103/PhysRevB.81.155210}.

\end{thebibliography}

\end{document}